\documentclass[12pt]{article}
\usepackage{amsmath,amsfonts,amssymb,amsxtra}
\usepackage{graphicx}
\input epsf

\usepackage[numbers,sort&compress]{natbib}

\makeatletter\@addtoreset{equation}{section}\makeatother

\renewcommand{\title}[1]{\vbox{\center\LARGE{#1}}\vspace{5mm}}
\renewcommand{\author}[1]{\vbox{\center#1}\vspace{5mm}}
\newcommand{\address}[1]{\vbox{\center\em#1}}
\newcommand{\email}[1]{\vbox{\center\tt#1}\vspace{5mm}}

\newcommand {\la} {\left \langle}
\newcommand {\ra} {\right \rangle}
\newcommand {\lb} {\left (}
\newcommand {\rb} {\right )}

\newcommand {\CalN} {\mathcal N}

\newcommand {\BR}   {\mathbb R}
\newcommand {\BZ}   {\mathbb Z}
\newcommand {\BC}   {\mathbb C}

\newcommand {\ep}  {\epsilon}

\newcommand{\g}{\mathfrak{g}}
\newcommand{\h}{\mathfrak{h}}

\newcommand {\p} {\partial}

\DeclareMathOperator{\tr} {tr}

\DeclareMathOperator{\vol}{vol}

\DeclareMathOperator{\Pexp} {Pexp}

\DeclareMathOperator{\diag}{diag}

\newcommand{\U}{U}

\newcommand{\const}{\mathrm{const}}

\numberwithin{equation}{section}

\begin{document}

\unitlength = .8mm

\bibliographystyle{utphys}

\begin{titlepage}
\begin{center}
\hfill{\tt ITEP-TH-17/09}\\
\hfill{\tt Imperial-TP-RR-01/2009}


\title{Notes on supersymmetric Wilson loops \\ on a two-sphere}

\author{Simone Giombi$^{1,a}$,
Vasily Pestun$^{1,*,b}$,
Riccardo Ricci$^{2,c}$}

\address{$^1$Center for the Fundamental Laws of Nature
\\Jefferson Physical Laboratory, Harvard University,\\
Cambridge, MA 02138 USA\\
\medskip
$^2$Theoretical Physics Group, Blackett Laboratory,\\
Imperial College, London, SW7 2AZ, U.K.}

\email{$^a$giombi@physics.harvard.edu,
$^b$pestun@physics.harvard.edu,
$^c$r.ricci@imperial.ac.uk}

\end{center}

\abstract{We study a recently discovered family of $1/8$-BPS supersymmetric Wilson
loops in ${\cal N} = 4$
super Yang-Mills theory and their string theory duals. The
operators are defined for arbitrary contours on a two-sphere in space-time, and
they were conjectured to be captured perturbatively by 2d bosonic Yang-Mills theory. In the $AdS$ dual, they are
described by
pseudo-holomorphic string surfaces living on a certain submanifold of
$AdS_5\times S^5$.
We show that the regularized area of these string surfaces is invariant under
area preserving diffeomorphisms of the boundary loop, in agreement with
the conjecture.
Further, we find a connection between the pseudo-holomorphicity
equations and an
auxiliary $\sigma$-model on $S^3$, which may help to construct new $1/8$-BPS string solutions. We also show that the conjectured relation to 2d Yang-Mills
implies that
a connected correlator of two Wilson loops is computed by a Hermitian
Gaussian two-matrix model. On the $AdS$ dual side, we argue that the
connected correlator is described by
two disconnected disks interacting through the exchange of supergravity modes,
and we show that this agrees with the strong coupling planar limit of
the two-matrix model.}

\vspace{0.9cm}
\flushbottom{$^* \overline{\text{On leave of absence from ITEP, 117259 Moscow, Russia.}}$}

\vfill

\end{titlepage}

\eject

\tableofcontents

\section{Introduction}

Exact results for non-protected operators in supersymmetric field theories which are valid for all values of the coupling constant $\lambda$ are typically beyond the reach of present techniques. For ${\cal N}=4$ super Yang-Mills with $SU(N)$ gauge group, AdS/CFT duality \cite{Maldacena:1998re,Gubser:1998bc,Witten:1998qj}, combined with the insights coming from integrability  \cite{Minahan:2002ve,Beisert:2003tq,Beisert:2004ry,Eden:2006rx,Beisert:2006ez,Freyhult:2007pz}, allows to obtain a wealth of non-trivial information on the spectrum of the theory. Nonetheless, exact results which interpolate between weak and strong coupling regimes are rare. 

A notable exception is the case of the half BPS circular Wilson loop \cite{Erickson:2000af,Drukker:2000rr,Semenoff:2001xp}. For this non-local operator an exact answer, valid for all values of $\lambda$ and $N$, is known. Indeed, it turns out that its expectation value is fully captured by a simple Hermitian matrix model with Gaussian potential. This remarkable result has been rigorously proved, using localization techniques, in \cite{Pestun:2007rz}. It is clearly important to generalize this example and find a richer class of observables which preserve less supersymmetry but still allow for an exact answer as this will provide new interpolating functions between weak and strong coupling regimes. In \cite{Drukker:2007dw,Drukker:2007qr} such a generalization was proposed by constructing a large new family of supersymmetric Wilson loop operators in the gauge theory. These operators are defined for arbitrary contours on a $S^3$ subspace of Euclidean spacetime $\mathbb{R}^4$, and couple to three of the six scalars of ${\cal N}=4$ super Yang-Mills in a suitable way which depends on the spacetime contour. Generically these loops preserve only two supercharges (more precisely, two combinations of Poincar\'e and superconformal supercharges). The half BPS circular loop is recovered as a special example when the contour is a great circle of $S^3$. 

Of particular interest is the restriction to the sub-sector of loops which lie on a great $S^2$ inside $S^3$. The restriction enhances the preserved supersymmetry to four supercharges. The dynamics of this sub-sector was conjectured to be fully captured by a reduced non-supersymmetric model, i.e. by bosonic 2d Yang-Mills on $S^2$, with a specific prescription which amounts to discarding the contribution of 2d instantons \cite{Drukker:2007yx,Drukker:2007qr}. According to this conjecture, the expectation value of the loops in ${\cal N}=4$ super Yang-Mills can be computed by studying the analogous observables in 2d Yang-Mills, and restricting to the ``zero-instanton sector''. As shown in \cite{Bassetto:2001mf,Bassetto:1999dg,Bassetto:1998sr}, this sector is again governed by the Hermitian Gaussian matrix model, with a simple rescaling of the coupling constant by a factor which depends on the area of the loop on $S^2$. In particular, for a loop along a great circle, one recovers the well known matrix model for the half BPS circle.   A first indication in support of the conjecture was given at weak coupling in \cite{Drukker:2007yx,Drukker:2007qr}, by observing that the combined ``gauge field+scalar'' propagator appearing in ladder diagrams effectively reduces to the propagator of pure 2d Yang-Mills in a suitable gauge with the WML prescription for the pole regularization. Further non-trivial perturbative evidence was presented at 2-loops in \cite{Bassetto:2008yf,Young:2008ed}. However, in \cite{Young:2008ed} a disagreement was found at order $\lambda^3$ for the connected correlator of two Wilson loops along two latitudes of $S^2$. 

In principle, it should be possible to prove (or disprove) the conjecture using localization of the ${\cal N}=4$ SYM path integral, in the same spirit of \cite{Pestun:2007rz}. A substantial step toward this goal has been taken in \cite{Pestun:2008thesis}\cite{Pestun:2009nn}, where it is shown that, due to the four supercharges preserved by the loops, the path integral indeed localizes on a 2d Higgs-Yang-Mills theory, which can be argued to be perturbatively equivalent to 2d Yang-Mills. However the computation of the associated one-loop determinant in the localization procedure appears to be technically challenging and was left open in \cite{Pestun:2008thesis}\cite{Pestun:2009nn}. In the case of the half BPS circular loop, the determinant was in fact shown to be trivial \cite{Pestun:2007rz}. If the same holds for the more general loops on $S^2$, the conjecture would be essentially proved. On the other hand, if the one-loop determinant is not trivial, it will deform the 2d theory with additional interactions which are not present in pure bosonic 2d Yang-Mills. All results we present in this paper fit naturally with the original form of the conjecture, hence trivial one-loop determinant, but they do not strictly speaking exclude the other possibility. A non-trivial one-loop determinant may explain the discrepancy found in \cite{Young:2008ed} for the connected correlator. However, we point out that such a deformation of the 2d theory, if present, is quite constrained by the available data: it should not contribute up to 2-loops in perturbation theory and should be irrelevant in the strong coupling limit for arbitrary loops (due to the results presented in this paper and in \cite{Drukker:2007yx,Drukker:2007qr}), and moreover it should not contribute {\it to all orders} for a loop at the equator, which is the half BPS circle for which the conjecture is rigorously proved \cite{Pestun:2007rz}. Another possible explanation for the disagreement found in \cite{Young:2008ed} is that there could be subtle issues in the choice of regularization scheme, which is not guaranteed to be compatible with the superconformal supercharges preserved by the loops. 

At strong coupling, Wilson loops in the fundamental representation\footnote{Wilson loops in higher rank representations can also be studied by replacing fundamental strings with D-branes with worldvolume electric flux, or more generally with the corresponding backreacted geometries, see {\it e.g.} \cite{Drukker:2005kx,Gomis:2006sb,Yamaguchi:2006tq},\cite{Giombi:2006de,Drukker:2006zk,
Okuda:2007kh,Okuda:2008px,Gomis:2008qa}.}  correspond to macroscopic strings in $AdS_5\times S^5$ ending on the loops at the holographic boundary \cite{Maldacena:1998im,Rey:1998ik}. Their expectation value is obtained by computing the  string action, after suitable regularization. From a purely mathematical point of view,  AdS/CFT duality therefore instructs us to find the minimal area worldsheet with the correct boundary conditions and to calculate the associated area. For a loop of arbitrary shape, this computation is prohibitive. On the other hand, the conjectured relation with 2d Yang-Mills suggests, if correct, a very simple answer which is valid for a wide class of Wilson loops. A preliminary confirmation of the 2d Yang-Mills conjecture at strong coupling was given in \cite{Drukker:2007yx,Drukker:2007qr} by studying the string configurations dual to a loop at a constant latitude angle and to a loop comprising two arcs of longitude.

One of the key-features of 2d Yang-Mills is the existence of a large symmetry group corresponding to the area preserving diffeomorphisms. To substantiate the 2d YM conjecture at strong coupling, we will prove that the regularized string action is indeed invariant under the action of this large symmetry group. To this end we will evaluate the variation of the on-shell string action as a function of the deformation of the boundary loop. In the calculation we will take advantage of the fact that the string worldsheets have a simple geometrical interpretation as surfaces calibrated by an almost complex structure $J$ of $AdS_4\times S^2$, and live on a $AdS_3\times S^2$ submanifold of the latter 6d space \cite{Drukker:2007qr}. The boundary of $AdS_3$ coincides with the two-sphere on which the gauge theory loops reside.

Supersymmetry usually drastically simplifies the understanding of string $\sigma$-models by reducing the dynamical problem to the search of solutions of a set of first order equations. For the class of loops here considered, the first order equations are the ``pseudo-holomorphic'' equations associated to $J$. Despite this great simplification, finding explicit solutions to these equations remains rather challenging. It is therefore of great help, as we show in this paper, that the solutions of the pseudo-holomorphic equations are automatically solutions of a much simpler model, {\it i.e.} of a non-linear $\sigma$-model on an auxiliary $S^3$.  If suitable boundary conditions are satisfied also the converse is true so that, given a solution of the auxiliary $\sigma$-model, the full string configuration in $AdS_3\times S^2$ can be reconstructed in a rather simple way. This approach offers therefore a new and insightful way to characterize the supersymmetric solutions of the string $\sigma$-model. For instance, by performing a reduction \`a la Pohlmeyer \cite{Pohlmeyer:1975nb}, known solutions can be reinterpreted as solitons of complex sinh-Gordon. We will be also able to prove that any ``stable'' solution to the pseudo-holomorphic equations is accompanied by an ``unstable'' one with equal and opposite regularized area, generalizing the analogous result for the case of the latitude loop considered in \cite{Drukker:2006ga}. This is in nice agreement with the strong coupling asymptotics of the Hermitian matrix model, where one finds two saddle points with equal and opposite exponents.

The auxiliary $\sigma$-model allows, in principle, to construct new solutions with greater ease. Indeed, as an application, we will find the string configuration which is dual to an infinitesimal deformation of the latitude solution. This ``wavy''-latitude solution can be obtained by studying the linearized perturbations around the latitude solution and provides another check of the 2d Yang-Mills at strong coupling. We also find a simple generalization of the latitude solution, which has the topology of a cylinder and physically corresponds to a subdominant contribution at strong coupling to the connected correlator of two coincident latitude loops with opposite orientations. 

We then move on to study the case of connected correlators of two distinct (non-intersecting) Wilson loops on $S^2$. We show that the structure of the pseudo-holomorphic equations implies that supersymmetric solutions with topology of a smooth cylinder connecting two distinct latitude loops do not exist. Building on this result, and by using a ``deformation'' argument, we will argue that the same conclusion holds for two non-coincident generic loops on $S^2$. This result implies that at strong coupling the connected correlator is described by a ``singular" cylinder consisting of two disconnected disks which interact by exchange of supergravity modes. 

In the second part of the paper, we test our string theory conclusions against the Hermitian Gaussian matrix model which follows from the conjectural relation to 2d Yang-Mills. In particular, we show that the restriction to the zero-instanton sector implies that correlators of arbitrary $1/8$ BPS Wilson loops on $S^2$ should be computed by an Hermitian Gaussian multi-matrix model. In general, exact results in 2d   
Yang-Mills with gauge group $U(N)$ can be written as unitary multi-matrix integrals \cite{Migdal:1975zf, Kazakov:1980zi, Kazakov:1981sb, Rusakov:1990rs, Blau:1993hj,Blau:1991mp, Gross:1989fc,Fine:1990zz,Witten:1991we,Gorsky:1993pe,Gross:1994ub}. As noted in \cite{Bassetto:2001mf}, discarding the contributions of the 2d unstable instantons amounts to neglecting the global structure of $U(N)$, i.e. to approximate the gauge group with its Lie algebra, namely the (anti)-hermitian matrices. For the case of the two-loop correlator, we explicitly solve at large $N$ the corresponding two-matrix model and obtain the strong coupling asymptotic expansion. The saddle points appearing at large $\lambda$ precisely agree with the bulk expectation of disconnected disk solutions. Of course, the two-matrix model predicts a concrete set of subleading corrections to the exponential saddle points. It would be important to check those corrections from the bulk, by computing the precise amplitude for the supergravity mode exchange (as well as quantum fluctuations of the disks themselves), especially in view of the disagreement found in \cite{Young:2008ed} for the connected correlator. 

The paper is organized as follows. In Section 2 we will provide a brief review of the pseudo-holomorphic equations which govern the dynamics of the loops at strong coupling. In Section 3 we will prove the invariance of the string action under area preserving diffeomorphisms at the boundary. In Section 4 we will derive the auxiliary $\sigma$-model on $S^3$ and discuss the associated Pohlmeyer reduction. In Section 5, we will construct the ``wavy''-latitude solution.  In Section 6 we will focus on the coincident latitude correlator, while correlators between generic non-coincident loops will be analyzed in Section 7. In Section 8 we discuss the matrix model computation for correlators of coincident loops, derive the Hermitian Gaussian multi-matrix model for correlators of non-coincident loops and solve explicitly the case of the two-matrix model.

\section{Review of pseudo-holomorphicity equations}

The Wilson loops introduced in \cite{Drukker:2007dw,Drukker:2007qr} are defined on a $S^3$ subspace of Euclidean spacetime $\mathbb{R}^4$ which we can take to be given by $x^{\mu}x^{\mu}=1$\footnote{By conformal invariance, we can take the boundary sphere to have radius 1. It is straightforward to restore the radius dependence if desired.}. They couple to three of the scalars of the ${\cal N}=4$ gauge multiplet as
\begin{equation}
 W({\cal C})=\tr \Pexp \oint_{\cal C} \left(A_\mu dx^\mu-i\, \frac{\sigma^i_R}{2}\Phi^i\right)\,,\qquad \mu=1,\ldots,4,\,\,i=1,2,3,
\label{S3-loops}
\end{equation}
where $\sigma^i_R\doteq 2\sigma^i_{\mu\nu}x^{\mu}dx^{\nu}$ are the left-invariant one-forms on $S^3$ (we follow \cite{Drukker:2007qr} in the conventions for the $SU(2)$ invariant one-forms)\footnote{An analogous construction in Minkowski spacetime, in which $S^3$ is replaced by the hyperbolic manifold $\mathbb{H}_3$, has been discussed in \cite{Branding:2009fw}.}. The coupling to the scalars is such that the loops are 1/16-BPS supersymmetric operators, preserving two combinations of Poincar\'e and superconformal supercharges. More details on the definition of the loops and on our conventions in the gauge theory will be given in Section 8. Here we will content ourselves with reviewing their dual string description.

Since the loops couple to the three of the scalars, it is natural to consider
an $AdS_5\times S^2$ subspace of $AdS_5\times S^5$. We write the metric on this subspace in the Poincar\'e patch as follows
\begin{equation}
\begin{aligned}
&ds^2=\frac{1}{z^2} dx^{\mu} dx^{\mu} + z^2 dy^i dy^i\,,\\
&z^{-2} \equiv y^i y^i\,.
\end{aligned}
\end{equation}
This is related to the standard metric in the direct product form by changing variables as $\theta^i=z y^i$, where $\theta^i$ is a unit vector parameterizing $S^2$.

It was shown in \cite{Drukker:2007qr} that the string worldsheets dual to the general Wilson loops on $S^3$ reside on the $AdS_4\times S^2$ subspace defined by
\begin{equation}
x^{\mu} x^{\mu} + z^2 = 1\,.
\label{Cons1}
\end{equation}
Note that the boundary of the $AdS_4$ is the $S^3$ on which the Wilson loop operators live. Furthermore it was shown in \cite{Drukker:2007qr} that the dual string worldsheets are pseudo-holomorphic surfaces with respect to an almost complex structure $J$ defined on this 6d subspace, i.e. the string embedding $X^M(\tau,\sigma)$ satisfy the equations
\begin{equation}
J^{M}_{~N} \partial_{\alpha}X^{N}=\sqrt{g}\epsilon_{\alpha\beta}\partial^{\beta}X^M\,,
\label{J-eq}
\end{equation}
where $M=1,\ldots,7$ and $X^M=(x^{\mu},y^i)$ are $AdS_5\times S^2$ embedding coordinates constrained according to (\ref{Cons1}). Here $\sqrt{g}$ is the determinant of the worldsheet metric, and our conventions for the antisymmetric $\epsilon$-symbol are such that $\epsilon_{\tau\sigma}=-1$ (we denote by $\tau$ the coordinate parameterizing the loop, and $\sigma$ the coordinate normal to the boundary). The explicit components of the almost complex structure $J$ read \cite{Drukker:2007qr}
\begin{equation}
J^{\mu}_{~\nu}=z^2\sigma_{\mu\nu}^i y^i,\,\,\,\,\,\,\,J^{\mu}_{~i}=z^2\sigma_{\mu\nu}^i x^{\nu}=-z^4 J^{i}_{~\mu},
\,\,\,\,\,\,\,J^i_{\ j}=-z^2 \epsilon_{ijk}y^k\,.
\label{J-components}
\end{equation}
 It is an interesting fact that the expression for $J$ is closely related, up to factors of $z$, to the almost complex structure of $S^6$ which is inherited from octonion multiplication. 

It is possible to explicitly show that solutions to (\ref{J-eq}) automatically satisfy the equations of motion and the Virasoro constraints for the $AdS_5\times S^5$ $\sigma$-model. The reduction of the dynamics to first order equations is of course a consequence of supersymmetry. In fact, by studying the $\kappa$-symmetry condition for the string, it can be shown that the pseudo-holomorphicity conditions (\ref{J-eq}) are equivalent to the requirement that the string solutions preserve the same set of supersymmetries of the dual gauge theory operators \cite{Drukker:2007qr}. 

Given the almost complex structure $J^M_{~N}$, it is natural to use the metric to define a corresponding two-form $J_{MN}$. It then follows that pseudo-holomorphic worldsheets are calibrated with respect to $J_{MN}$: their area, or equivalently their on-shell action $S$, is given by
\begin{equation}
S=\frac{\sqrt{\lambda}}{2\pi}\int_{\Sigma} J\,.
\end{equation}  
Note that the two-form $J$ is not closed\footnote{The non-closure of $J$ points to a similarity with the case of a ``generalized calibration", see e.g. \cite{Gutowski:1999iu}.} and it is therefore not straightforward to obtain a general expression for the worldsheet area in terms of boundary data. However, it was shown in \cite{Drukker:2007qr} that $J$ admits a splitting
\begin{equation}
J=J_0+d\Omega\,.
\label{J-split}
\end{equation}
where $\Omega$ can be chosen in such a way that the exact piece $d\Omega$ cancels the usual boundary term needed to regularize the infinite worldsheet area, namely the ``Legendre transform'' over the $AdS$ radial coordinate $z$ \cite{Drukker:1999zq} 
\begin{equation}
\int_{\partial \Sigma} \Omega = \int_{\partial \Sigma}d\tau\,z P_z\,.
\end{equation}
Therefore the regularized action $S_{reg}=S-\int_{\partial \Sigma}d\tau\,z P_z$ can be written as
\begin{equation}
S_{reg}=\frac{\sqrt{\lambda}}{2\pi}\int_{\Sigma}J_0=\frac{\sqrt{\lambda}}{4\pi}\int d^2\sigma\sqrt{g}\left(\partial_{\alpha}\theta^i \partial^{\alpha}\theta^i+\frac{1}{z}\nabla^2z\right)\,.
\label{Sreg}
\end{equation}

In this paper we will be interested in the special subclass of 1/8-BPS Wilson loops living on a great $S^2\subset S^3$. The corresponding string solutions are obtained from the above general construction by restricting to the $AdS_3 \times S^2$ subspace given by $x^4=0$.
It is then natural to use a notation for the coordinates such that $x$ and $y$ carry the same 3d-vector index $(x^{\mu},y^i)\rightarrow(x^i,y^i)$. Accordingly, the components of the left-invariant one form simplify as $\sigma^i_{\mu\nu}\rightarrow\epsilon_{ijk}$. All pseudo-holomorphicity conditions in (\ref{J-eq}) still hold, but the $M=4$ component simplifies to
\begin{equation}
\partial_{\alpha}\left(x^iy^i\right)=0\,,
\end{equation}
which is readily solved by 
\begin{equation}
x^i y^i = C\,,
\label{Cons2}
\end{equation}
where $C$ is an integration constant. 

\section{Deformations of $1/8$-BPS Wilson loops and invariance under area preserving diffeomorphisms}
\label{deform-section}
Consider an arbitrary curve ${\mathcal C}$ on $S^2$ corresponding to a $1/8$-BPS Wilson loop operator. For any such curve, let us assume that we have found a corresponding dual string worldsheet $X^M(\tau,\sigma)$ which solves the pseudo-holomorphicity conditions (\ref{J-eq})\footnote{Though this is a reasonable assumption, we were not able to prove in general that solutions of (\ref{J-eq}) exist for any given boundary loop ${\mathcal C}$. An indirect argument based on holography is that, since in the gauge theory supersymmetric operators exist for arbitrary curves, we expect corresponding dual string solutions preserving the same supersymmetry to also exist.}. Now consider an arbitrary small perturbation of the boundary loop ${\mathcal C}\rightarrow{\mathcal C}+\delta{\mathcal C}$ such that the perturbed loop still sits inside $S^2$. Correspondingly, the string solution in the bulk will be deformed to a new solution of the pseudo-holomorphicity equations 
\begin{equation}
X^M(\tau,\sigma) \rightarrow X^M(\tau,\sigma)+\delta X^M(\tau,\sigma)\,.
\label{X-var}
\end{equation}
For convenience, we will assume throughout that the solutions are in the conformal gauge $\sqrt{g}g^{\alpha\beta}=\delta^{\alpha\beta}$, and to be concrete we can imagine that $0<\tau<2\pi$ is a periodic coordinate parameterizing the loop ${\mathcal C}$ and $0<\sigma<\infty$ is the coordinate normal to the boundary, with the loop ${\mathcal C}$ sitting at $\sigma=0$.   

We would now like to determine how the on-shell string action varies as a function of the deformation of the boundary loop. In conformal gauge, the Polyakov action reads (for simplicity we set $\sqrt{\lambda}/2\pi=1$)
\begin{equation}
S=\frac{1}{2}\int_{\Sigma} d^2\sigma \partial_{\alpha}X^M\partial_{\alpha}X^N G_{MN}\,.
\end{equation}
It is easy to see that the first order variation of the action resulting from (\ref{X-var}) is a boundary term, since $X^M$ is a solution of the equations of motion
\begin{equation}
\delta S = - \int_{{\partial \Sigma}} d\tau\, \delta X^M \partial_{\sigma}X^N G_{MN}\,.
\end{equation} 
Now we can use (\ref{J-eq}) to rewrite this as
\begin{equation}
\delta S = \int_{{\partial \Sigma}} d\tau\, \delta X^M \partial_{\tau}X^N J_{MN}\,.
\label{bterm}
\end{equation}
The integrand has to be evaluated in the limit $\sigma \rightarrow 0$, so one needs to know the behaviour of the solution and of the perturbation close to $\sigma=0$. On general grounds one expects an expansion of the form\footnote{We assume that the $AdS_3\times S^2$ coordinates $\left(x^i(\tau,\sigma),z(\tau,\sigma),\theta^i(\tau,\sigma)\right)$ are analytic functions at $\sigma=0$, i.e. they admit a Taylor expansion in positive powers of $\sigma$. This is true for all known explicit solutions, though we do not have a first-principle prove of this assumption. The fact that $y^i(\tau,\sigma)$ starts as $\frac{1}{\sigma}$ is just due to the definition $y^i=\frac{\theta^i}{z}$ and the fact that $z$ vanishes at the boundary.}  
\begin{equation}
\begin{aligned}
&x^i(\tau,\sigma)=x^i_{(0)}(\tau)+\sigma x^i_{(1)}(\tau)+\sigma^2 x^i_{(2)}(\tau)+\ldots \\
&y^i(\tau,\sigma)=\frac{y^i_{(0)}(\tau)}{\sigma}+y^i_{(1)}(\tau)+\sigma y^i_{(2)}(\tau)+\ldots\,.
\label{expansion}
\end{aligned}
\end{equation}
Here $x^i_{(0)}(\tau)$ parameterize the curve ${\mathcal C}$ on $S^2$, and $y^i_{(0)}(\tau)$ will be related to the couplings $\theta^i(\tau)$ to the scalar fields in the $1/8$-BPS Wilson loop operator, see below.

Plugging into (\ref{bterm}) the explicit components of $J_{MN}$ (which can be obtained from (\ref{J-components}) by lowering an index with the metric)
one can see that the $J_{ij}$ component does not contribute in the $\sigma \rightarrow 0$ limit and the first order variation of the action can be written as 
\begin{equation}
\delta S = -\int_{{\partial \Sigma}} d\tau \left(\delta x^i \partial_{\tau}(y^j x^k)+\delta y^i \partial_{\tau}x^j x^k\right)\epsilon_{ijk}\,.
\label{bterm-simple}
\end{equation}
By inspecting the expansion (\ref{expansion}) one can see that only the first two non-trivial orders of $X$ and $\delta X$ are needed in order to determine $\delta S$. These can be obtained by solving perturbatively in $\sigma$ the pseudo-holomorphicity conditions and the constraints (\ref{Cons1}) and (\ref{Cons2}). It is easy to see from the pseudo-holomorphicity conditions with $M=1,2,3$ that 
\begin{equation}
 x_{(1)}^i=\delta x_{(1)}^i=0\,,
\end{equation}
so the variation of the action is given by
\begin{equation}
\begin{aligned}
\delta S=&- \frac{1}{\sigma}  \int_{\partial \Sigma} d\tau \left(\delta x^i_{(0)} \partial_{\tau}(y^j_{(0)} x^k_{(0)})+\delta y^i_{(0)} \dot x^j_{(0)} x^k_{(0)}\right)\epsilon_{ijk} \\
&-\int_{\partial \Sigma} d\tau \left(\delta x^i_{(0)} \partial_{\tau}(y^j_{(1)} x^k_{(0)})+\delta y^i_{(1)}\dot x^j_{(0)} x^k_{(0)}\right)\epsilon_{ijk}+\mathcal{O}(\sigma)\,.
\label{deltaS}
\end{aligned}
\end{equation}
The leading two orders of $y^i$ and $\delta y^i$ can be determined from the first two orders of the pseudo-holomorphicity conditions in the directions $M=5,6,7$ and of the constraint (\ref{Cons2}). The leading order gives the condition
\begin{equation}
y^i_{(0)}=\epsilon_{ijk}\frac{x_{(0)}^j\dot{x}_{(0)}^k}{|\dot{x}_{(0)}|^2}\,,
\label{y0}
\end{equation}
which is nothing but the scalar coupling condition for the 1/8-BPS Wilson loop operators \cite{Drukker:2007yx}, see the definition of the gauge theory operators in eq. (\ref{eq:define-loop}) (to obtain the above relation, we used the fact that $|\dot x_{(0)}| = |\partial_{\sigma}z|_{\sigma \rightarrow 0}$, which generally holds for string worldsheets dual to Wilson loops \cite{Drukker:1999zq}, and can be also seen to follow in the present case from the $\sigma$-expansion of (\ref{J-eq}) and (\ref{Cons1})). Varying equation (\ref{y0}) one then gets the corresponding expression for $\delta y^i_{(0)}$.  Plugging into the first line of (\ref{deltaS}) one can see that it vanishes for arbitrary loop, so there is no divergent piece in $\delta S$. Consistently with this fact, it is also possible to check that the boundary term which has to be added to the action in order to remove divergences, i.e. $-\int_{\partial \Sigma} z P_z$, does not give contributions at first order in $\delta X$. Hence we have that $\delta S=\delta S_{reg}$ (see also the discussion at the end of Section \ref{S3-section}). 

At next order one gets the relations
\begin{equation}
\begin{aligned}
&\epsilon_{ijk} x^j_{(0)} \dot x^k_{(0)} y^i_{(1)}=0\\ 
&x^i_{(0)}y^i_{(1)}=C\,,
\end{aligned}
\end{equation}
where $C$ is the integration constant in (\ref{Cons2}). These are solved by
\begin{equation}
y^i_{(1)}=C x^i_{(0)}+a(\tau) \dot x^i_{(0)}\,,
\label{y1-sol}
\end{equation}
where $a(\tau)$ is an undetermined function, and correspondingly for the fluctuations
\begin{equation}
\delta y^i_{(1)}=C \delta x^i_{(0)} + a(\tau) \delta\dot x^i_{(0)}+\delta C x^i_{(0)} + \delta a(\tau) \dot x^i_{(0)}\,.
\end{equation}
Fortunately, the undetermined function $a(\tau)$ as well as $\delta C$ and $\delta a$ turn out to be not needed in the evaluation of $\delta S$. Plugging the above solutions into $\delta S$ and integrating by parts the first term in the second line of (\ref{deltaS}), one finally obtains 
\begin{equation}
\delta S= -C \int_{{\mathcal C}} d\tau \delta x^i_{(0)} \dot x^j_{(0)} x^k_{(0)} \epsilon_{ijk}\,.
\end{equation}
Now one can recognize that the integral is precisely the first order variation of the area $A_1$ of the loop $\mathcal C$ on $S^2$ (we define $A_1$ to be the area of the region of $S^2$ bounded by the loop and including the north pole). One easy way to see this is for example to express the loop in polar coordinates $(\theta(\tau),\phi(\tau))$ where the expression for the area and its variation are 
\begin{equation}
\begin{aligned}
&A_1=\int_{\mathcal C} d\tau \dot\phi\left(1-\cos\theta\right)\\
&\delta A_1 = \int_{\mathcal C} d\tau \sin\theta\left(\dot \phi \delta \theta- \dot\theta \delta\phi\right)\,.
\end{aligned}
\end{equation}  
Going to cartesian coordinates one can see that indeed
\begin{equation}
\delta A_1=\int_{{\mathcal C}} d\tau \delta x^i_{(0)} \dot x^j_{(0)} x^k_{(0)} \epsilon_{ijk}\,,
\end{equation}
so our final result for the variation of the action is 
\begin{equation}
\delta S=-C\delta A_1\,.
\label{S-variation}
\end{equation}
This proves that the string action is invariant under an arbitrary deformation of the boundary loop which preserves its area. Since the above deformation argument was carried out for arbitrary loop, it follows that  any two loops with the same area will correspond to string worldsheets with the same value of the on-shell action. To fix the general functional form of the string action in terms of the loop area it is then sufficient to know its value for a simple class of loops, like the 1/4-BPS latitudes of \cite{Drukker:2006ga}\cite{Drukker:2007qr}, see eq. (\ref{latitude}) below, from which one infers (we reinsert here the explicit $\lambda$ dependence)
\begin{equation}
S_{reg}=-\frac{\sqrt{\lambda}}{2\pi}\sqrt{A_1A_2}\,,
\label{area-formula}
\end{equation}
where $A_2=A-A_1$ ($A=4\pi$ is the total area of the sphere), in agreement with the conjecture that these loops are captured by the perturbative sector of 2d Yang-Mills theory on $S^2$, or equivalently by the Gaussian Hermitian matrix model with a rescaled `t Hooft coupling $\lambda \rightarrow \frac{A_1A_2}{4\pi^2}\lambda$, see Section \ref{Matrix-section}. Note that in (\ref{area-formula}) we have written the result corresponding to the dominant contribution at strong coupling. It was pointed out in \cite{Drukker:2006ga} in the case of the $1/4$-BPS latitudes that for the same loop there are two supersymmetric solutions, one ``stable'' and one ``unstable'', which have equal and opposite regularized areas: these correspond to whether the worldsheet wraps the smaller/larger region of the ``dual'' $S^2\subset S^5$ with respect to the scalar coupling loop $\theta^i(\tau)$. Intuitively one would expect this to be a general feature of arbitrary $1/8$-BPS loops. In fact, in Section \ref{unstable} we will give a simple argument that shows that for every given loop on the boundary $S^2$, there are two solutions to the pseudo-holomorphicity equations, which are related by $C \rightarrow -C$. From eq. (\ref{S-variation}) one then sees that the two solutions should have equal and opposite regularized action. It is remarkable that both stable and unstable solutions appear as saddle points in the large $\lambda$ expansion of the Gaussian matrix model, as will be reviewed later.

Note that, viewing (\ref{S-variation}) as a differential equation for $S$ whose solution is (\ref{area-formula}), the above calculation implies that the constant $C$ does not depend on the shape of the loop and takes the value
\begin{equation}
C=\pm\frac{A_2-A_1}{2\sqrt{A_1A_2}}\,,
\label{CC}
\end{equation}
where the $+/-$ correspond to stable/unstable worldsheet as explained above. It would be nice to prove this relation by independent means. We have not been able to do this for the general case, but in the next subsection we will show that (\ref{CC}) can be proven to be correct in the limit of very small loops. Before doing that, we can show as a check that (\ref{CC}) agrees with the known solutions. For example, by looking at the explicit latitude solution \cite{Drukker:2006ga}\cite{Drukker:2007qr} (the choice of $\pm$ sign corresponds to the stable/unstable solutions)
\begin{equation}
\begin{aligned}
&x^1=\frac{\tanh\sigma_0 \cos\tau}{\cosh\sigma},\,\qquad x^2=\frac{\tanh\sigma_0 \sin\tau}{\cosh\sigma},\qquad x^3=\frac{1}{\cosh\sigma_0}\\
&y^1=-\frac{\cos\tau}{z\cosh(\sigma_0\pm\sigma)},\,\qquad y^2=-\frac{\sin\tau}{z\cosh(\sigma_0\pm\sigma)},\,\qquad
y^3=\frac{\tanh(\sigma_0\pm\sigma)}{z}\,\\
&z=\tanh\sigma_0\tanh\sigma\,,
\label{latitude}
\end{aligned}
\end{equation}
and recalling that the parameter $\sigma_0$ is related to the latitude angle by $\tanh\sigma_0=\sin\theta_0$, one gets indeed $C=\pm\cot\theta_0$, which agrees with (\ref{CC}) since $A_{1,2}=2\pi\left(1\mp\cos\theta_0\right)$. Another example of explicit solution is the $1/4$-BPS string corresponding to a loop made of two half longitudes with an opening angle $\delta$ \cite{Drukker:2007qr}. In conformal gauge, the solution can be written as\footnote{Here we write the ``stable'' solution. Analogously to the case of the latitude, there is a corresponding unstable counterpart, see Section \ref{S3-section}.}
\begin{equation}
\begin{aligned}
&x^1=\frac{a \sin a\sigma\sin\sigma+\cos a\sigma \cos \sigma}{\cosh\sqrt{1-a^2}\tau}\,, \qquad 
x^2=\frac{a \cos a\sigma\sin\sigma-\sin a\sigma \cos \sigma}{\cosh\sqrt{1-a^2}\tau}\,,\\
&x^3=-\tanh\sqrt{1-a^2}\tau\,,\qquad \qquad z=\frac{\sqrt{1-a^2}\sin\sigma}{\cosh\sqrt{1-a^2}\tau}\,,\\
&y^1=\frac{\sin a \sigma}{z}\,,\qquad \qquad y^2 = \frac{\cos a\sigma}{z} \qquad \qquad
y^3=0\,.
\label{longitudes}
\end{aligned}
\end{equation}
where the parameter $a$ is related to the opening angle by $a=\frac{\pi-\delta}{\pi}$, and the range of the coordinates is $-\infty<\tau<\infty$, $0<\sigma<\pi$. From this solution we get $C=x^iy^i=\frac{a}{\sqrt{1-a^2}}$, again in agreement with (\ref{CC}) after using $A_1=2\delta$ and $A_2=2(2\pi-\delta)$.

As a final remark, notice that the simplicity of the end result (\ref{S-variation}) crucially depends on the structure of the pseudo-holomorphic equations in the special $1/8$-BPS case. In the more general case of $1/16$-BPS loops on $S^3$, one could carry on a similar deformation argument with the result
\begin{equation}
\delta S=-\int_{\partial \Sigma} d\tau  y^i_{(1)} \delta x_{(0)}^\mu \dot{x}_{(0)}^\nu \sigma^i_{\mu \nu}\,.
\end{equation}
Unfortunately in this case we cannot easily determine $y^i_{(1)}$ as done in (\ref{y1-sol}) for the $1/8$-BPS case, mainly because of the absence of the simple constraint $x^i y^i = C$.

\subsection{The flat space limit}
Consider the limit in which the loop is very small compared to the radius of the sphere. Then the curvature of the sphere can be neglected and one ends up with a loop on flat space. By studying the supersymmetry preserved by the corresponding gauge theory operators, it was shown in \cite{Drukker:2007qr}\cite{Drukker:2007dw} that in this limit one recovers the Wilson loops of Zarembo \cite{Zarembo:2002an} which preserve some fraction of Poincar\'e supersymmetry. Correspondingly, we expect that eq. (\ref{J-eq}) should go in this limit to the pseudo-holomorphicity equations of \cite{Dymarsky:2006ve}, which describe Zarembo's loop in the bulk. Specializing to the case of loops on $S^2$\footnote{The case of loops on $S^3$ should go exactly in the same way, and one will recover the equations of \cite{Dymarsky:2006ve} for supersymmetric Wilson loops on $\mathbb{R}^3$. However the case of $1/16$-BPS supersymmetric Wilson loops on $\mathbb{R}^4$ cannot evidently be recovered as a limit of our loops on $S^3$.}, this limit can be systematically defined by setting
\begin{equation}
\begin{aligned}
&x^1=\epsilon\, \hat{x}^1\,,\qquad x^2=\epsilon\, \hat{x}^2\,,\qquad z=\epsilon\, \hat{z}\,,\\
&y^i=\frac{1}{\epsilon}\hat{y}^i\,,\quad i=1,2,3\,,
\label{flat-limit}
\end{aligned}
\end{equation}
and taking the limit $\epsilon\rightarrow 0$ while keeping the hatted coordinates fixed (note that $x^3$ can be expressed in terms of $x^1$,$\,x^2$ and $z$ using (\ref{Cons1})). The pseudo-holomorphicity equations go in this limit to 
\begin{equation}
\begin{aligned}
&\partial_{\alpha}\hat{x}^1=\hat{z}^2 \epsilon_{\alpha\beta}\partial_{\beta}\hat{y}^2\,,\qquad
\qquad\partial_{\alpha}\hat{x}^2=-\hat{z}^2 \epsilon_{\alpha\beta}\partial_{\beta}\hat{y}^1\,,\\
&\partial_{\alpha}\hat{y}^3=0\,,
\label{Jeq-flat}
\end{aligned}
\end{equation}
which indeed are the same as the equations of \cite{Dymarsky:2006ve} for the case of loops on $\mathbb{R}^2$ (up to a trivial renaming of the $\hat{y}$ variables). The equation on the second line implies $\hat{y}^3=\hat{c}$, where $\hat{c}$ is related to $C$ in eq. (\ref{Cons2}) by $C=\hat{c}/\epsilon$. It is convenient to introduce coordinates on $S^2$ as 
\begin{equation}
\hat{\theta}^i=\hat{z}\left(\hat{y}^2,-\hat{y}^1,\hat{y}^3\right)\,,
\end{equation}
so that eq. (\ref{Jeq-flat}) can be written as
\begin{equation}
\begin{aligned}
&\partial_{\alpha}\hat{x}^i=\frac{1}{\hat{c}}\epsilon_{\alpha\beta} j^{3i}_{\beta}\,,\\
&\hat{\theta}^3=\hat{c}\hat{z}\,,
\end{aligned}
\label{Jflat-S2}
\end{equation}
where $j^{3i}_{\alpha}=\hat{\theta}^3 \partial_{\alpha} \hat{\theta}^i-\hat{\theta}^i \partial_{\alpha} \hat{\theta}^3$ are components of the $SO(3)$ current. This way of writing the equations displays a connection to an auxiliary $S^2$ $\sigma$-model for the variables $\hat{\theta}^i$, and this was used in \cite{Dymarsky:2006ve} to show that the constant $\hat{c}$ only depends on the area of the loop. This can be seen in the following way. The area of the loop $\hat{x}$ on $\mathbb{R}^2$ is given by
\begin{equation}
\begin{aligned}
A_{2d}=-\int d^2\sigma \epsilon_{\alpha\beta} \partial_{\alpha}\hat{x}^1\partial_{\beta}\hat{x}^2=
\frac{1}{\hat{c}^2}\int d^2\sigma\epsilon_{\alpha\beta} j^{13}_{\alpha} j^{32}_{\beta}\,.
\end{aligned}
\end{equation}
Using an explicit parameterization $\hat{\theta}^i=\left(\sin v\cos w,\sin v \sin w,\cos v\right)$ this becomes
\begin{equation}
A_{2d}=\frac{1}{\hat{c}^2}\int dw dv  \sin v \cos v = \frac{\pi}{\hat{c}^2}\,,
\label{A2d}
\end{equation} 
where in the last equality we have used that the worldsheet wraps half of the sphere (since $\hat{z}>0$). Since in the limit (\ref{flat-limit}) the areas $A_1,\,A_2$ defined before goes to 
\begin{equation}
A_1 \rightarrow \epsilon^2 A_{2d}\,,\qquad A_2 \rightarrow 4\pi\,,
\end{equation}
the formula (\ref{A2d}) implies that in this limit we have
\begin{equation}
C^2=\frac{\pi}{A_1}\,,
\end{equation}
in agreement with the expected general relation (\ref{CC}).

The relation to the auxiliary $S^2$ $\sigma$-model in (\ref{Jflat-S2}) is a special case of a more general relation to a $S^4$ $\sigma$-model valid for the general Zarembo loops on $\mathbb{R}^4$ \cite{Dymarsky:2006ve}. A similar connection to a $\sigma$-model on an auxiliary $S^3$ holds for the case of our 1/8-BPS supersymmetric Wilson loops, as will be discussed in the next section.  

\section{Pseudo-holomorphicity equations and $S^3$ $\sigma$-model}
\label{S3-section}
In this section we show that, in the case of $1/8$-BPS loops, there exist a correspondence between solutions to the pseudo-holomorphicity equations (\ref{J-eq})
and solutions of an auxiliary $S^3$ $\sigma$-model. This may be useful to better understand general properties of these supersymmetric worldsheets and hopefully to find new solutions. 

For convenience, we will work again in the conformal gauge $\sqrt{g}g^{\alpha\beta}=\delta^{\alpha\beta}$. Using the explicit form of $J$ in (\ref{J-components}) specialized to the case $x^4=0$, the full set of pseudo-holomorphic equations corresponding to the $1/8$-BPS loops on $S^2$ can be conveniently written in vector notation as 
\begin{equation}
\begin{aligned}
&z^2\partial_{\alpha}\left(\vec{x}\times \vec{y}\right)=\epsilon_{\alpha\beta}\partial_{\beta}\vec{x}\,,\\
&\vec{x}\cdot\vec{y}=C\,,\\
&\frac{1}{z^2}\vec{x}\times\partial_{\alpha}\vec{x}+z^2\vec{y}\times\partial_{\alpha}\vec{y}=
\epsilon_{\alpha\beta}\partial_{\beta}\vec{y}\,. 
\end{aligned}
\label{equations-explicit}
\end{equation} 

Here the first two lines correspond to the $M=\mu$ component of (\ref{J-eq}) and the last line to $M=i$. Note that these equations are not all independent. In fact, while (\ref{J-eq}) are a set of fourteen equations, only at most seven can be independent, since $J^2=-1$. So by this counting argument, one may expect that once the first two lines of (\ref{equations-explicit}) are solved, the remaining three equations should be automatically satisfied. This can be seen as follows. Define
\begin{equation}
\begin{aligned}
&v_{\alpha}^{\mu}=J^{\mu}_{~N} \partial_{\alpha}X^{N}-\epsilon_{\alpha\beta}\partial_{\beta}x^{\mu}\\
&w_{\alpha}^i=J^{i}_{~N} \partial_{\alpha}X^{N}-\epsilon_{\alpha\beta}\partial_{\beta}y^{i}\,,
\end{aligned}
\label{arg1}
\end{equation}
so that $v_{\alpha}^{\mu}=0$ gives the first two lines of (\ref{equations-explicit}) and $w_{\alpha}^i=0$ the last one. Using $J^2=-1$ one can show that
\begin{equation}
J^{\mu}_{~i} w^i_{\alpha}=-J^{\mu}_{~\nu}v_{\alpha}^{\nu}-\epsilon_{\alpha\beta}v_{\beta}^{\mu}\,,
\end{equation} 
so if the first two lines of (\ref{equations-explicit}) are satisfied one gets $J^{\mu}_{~i} w^i_{\alpha}=0$. Separating out the $\mu=4$ component and using the explicit form of $J$ this gives
\begin{equation}
x^{\mu}\sigma^i_{\mu j} w_{\alpha}^i =0\,,\qquad \qquad  x^i w_{\alpha}^i=0\,.
\label{w-eq}
\end{equation}
The first equation can have non-trivial solutions if the determinant of $x^{\mu}\sigma^i_{\mu j}$ vanishes. If $x^4=0$, corresponding to $1/8$-BPS loops, this in fact happens, and the zero-eigenvector is proportional to $x^i$. So the first equation (\ref{w-eq}) is solved by $w_{\alpha}^i=c_{\alpha}x^i$, with $c_{\alpha}$ some arbitrary function. But the second equation would then imply $c_{\alpha}=0$, hence the only solution is $w_{\alpha}^i=0$. 

This argument shows that it is sufficient to concentrate on the first two lines of (\ref{equations-explicit}) in order to solve the pseudo-holomorphicity conditions. The first equation implies
\begin{equation}
\partial_{\alpha}\left(z^2\partial_{\alpha}(\vec{y}\times\vec{x})\right)=0\,.
\label{2ndorder}
\end{equation}
Now define a 4-vector $\xi^A=(\vec{\xi},\xi^4)$ as 
\begin{equation}
\vec{\xi}=z\, \vec{y}\times \vec{x}\,,\qquad \qquad \xi^4=\sqrt{1+C^2}z\,.
\label{xiA}
\end{equation} 
Using the fact that 
\begin{equation}
\left(\vec{y}\times\vec{x}\right)^2=y^2 x^2-(\vec{x}\cdot \vec{y})^2=\frac{1}{z^2}-1-C^2\,,
\end{equation}
where in the second equality we have used (\ref{Cons1}) and (\ref{Cons2}), it is easy to show that
\begin{equation}
\xi^A\xi^A=1\,,
\end{equation}
so $\xi^A$ are coordinates on a round three-sphere. In terms of the $\xi^A$, eq. (\ref{2ndorder}) becomes
\begin{equation}
\xi^4 \partial^2 \xi^i - \xi^i \partial^2 \xi^4=0\,.
\end{equation}
Multiplying this equation with $\xi^i$ and using $\xi^A\xi^A=1$, it can be seen to be equivalent to 
\begin{equation}
\partial^2\xi^A+\partial_{\alpha}\xi^B\partial_{\alpha}\xi^B\,\xi^A=0\,.
\label{eq-S3-smodel}
\end{equation} 
These are precisely the equations of motion of a $S^3$ $\sigma$-model with action 
\begin{equation}
S=\frac{1}{2}\int d^2\sigma \left(\partial_{\alpha}\xi^A\partial_{\alpha}\xi^A +\Lambda(\xi^A\xi^A-1)\right)\,,
\label{S3-action}
\end{equation}
where we have introduced a Lagrange multiplier that enforces the condition $\xi^A\xi^A=1$. 

Note that since $z>0$, the solution is constrained to lie in the upper hemisphere of the $S^3$. It is not difficult to work out the appropriate boundary conditions at $\sigma \rightarrow 0$. First of all $z$, and so $\xi^4$, has to vanish at $\sigma=0$, and then by expanding the 3-vector $\xi^i$ close to $\sigma=0$ in terms of $x^i$ and $y^i$, one concludes that the boundary conditions are
\begin{equation}
\xi^4 |_{\sigma=0}=0 \,,\qquad \qquad \xi^i |_{\sigma=0} = \frac{\dot x_{(0)}^i}{|\dot x_{(0)}|}\,.
\label{S3-bc}
\end{equation}
So the appropriate solution is a worldsheet ending on the $S^2$ equator of this auxiliary $S^3$, where it describes a loop parameterized by the unit tangent vector to the field theory loop $x_{(0)}^i$.

So we have shown that to every solution to the pseudo-holomorphicity equations there corresponds a solution to an auxiliary $S^3$ $\sigma$-model with boundary conditions (\ref{S3-bc}). Clearly one would like to use this relation in the opposite direction and map the problem of solving the pseudo-holomorphicity conditions to the problem of finding classical solutions to the $S^3$ $\sigma$-model. In principle, once a solution to (\ref{eq-S3-smodel}) with the appropriate boundary conditions is given, the full solution $x^i,y^i$ may be be reconstructed as follows. First, the $x^i$ can be determined by solving the elementary first order differential equation, which follows from the first line of (\ref{equations-explicit})
\begin{equation}
\partial_{\alpha}x^i =  \frac{1}{\sqrt{1+C^2}}\epsilon_{\alpha\beta}\left(\xi^4\partial_{\beta}\xi^i-\xi^i \partial_{\beta}\xi^4\right)\,.
\label{x-sol2}
\end{equation}
As a remark, note that the right-hand side of this equation is proportional to a component of the conserved $SO(4)$ current $j^{AB}_{\alpha}=\xi^A\partial_{\alpha}\xi^B-\xi^B\partial_{\alpha}\xi^A$. Once $x^i$ are known, the $y^i$ can be determined algebraically using the identity
\begin{equation}
\vec{x}\times\vec{y}\times\vec{x}=(1-z^2) \vec{y}-C \vec{x}\,,
\end{equation}
which gives 
\begin{equation}
\vec{y}=\frac{1}{1-z^2}\left(\frac{1}{z} \vec{x}\times\vec{\xi}+C \vec{x}\right)\,,
\label{y-sol}
\end{equation}
with $z=\xi^4/\sqrt{1+C^2}$. The last line of (\ref{equations-explicit}) will then be automatically satisfied following the argument from (\ref{arg1}) to (\ref{w-eq}).

In practice, however, it is not straightforward to apply successfully this algorithm because it is not easy to keep under control the boundary conditions on $AdS_3\times S^2$ from the point of view of the auxiliary $S^3$ $\sigma$-model. In particular, once a solution on $S^3$ is given, it is not guaranteed that after integrating (\ref{x-sol2}) the resulting solution will satisfy the constraint $x^2+z^2=1$. Besides this, there is also a self-consistency condition of the reduction to $S^3$, namely $\xi^ix^i=0$ (see the definition eq. (\ref{xiA})), which is not guaranteed to hold in general from integrating (\ref{x-sol2}). It would be nice to understand if there is an efficient way to implement those constraints directly at the level of the $S^3$ variables.

As an example for which the algorithm can be efficiently applied, consider the following simple solution of the $S^3$ $\sigma$-model
\begin{equation}
\xi^4 = \tanh \sigma\,,\qquad \qquad \xi^i=\left(-\frac{\sin \tau}{\cosh\sigma},\frac{\cos\tau}{\cosh\sigma},0\right)\,.
\label{S3-latitude}
\end{equation} 
This is a minimal surface inside $S^3$ which ends along a great circle of the equatorial $S^2$. Geometrically, it is itself half of a great two-sphere inside $S^3$, as can be seen from the induced metric
\begin{equation}
ds^2=\frac{1}{\cosh^2\sigma}\left(d\tau^2+d\sigma^2\right)\,.
\label{half-S2}
\end{equation}
Solving (\ref{x-sol2}) and (\ref{y-sol}) one easily finds that this solution generates the $1/4$-BPS latitude in (\ref{latitude}). Note that, consistently with the fact that the unit tangent vector to a latitude is independent from the latitude angle, all $1/4$-BPS latitudes correspond to the same solution on $S^3$, the difference between them only comes from the choice of the integration constant $C$ in (\ref{x-sol2}). This constant is fixed by imposing the constraint $x^2+z^2=1$, up to an overall sign choice. This sign ambiguity only affects the $y^i$, see eq. (\ref{y-sol}), and corresponds to the stable and unstable solutions in (\ref{latitude}). The existence of stable and unstable worldsheets for each loop is a general fact which follows from the structure of the pseudo-holomorphicity equations, as will be explained in Section \ref{unstable}.

As a remark, notice that not all solutions of the auxiliary $S^3$ $\sigma$-model have an interpretation as minimal surfaces. This is only true if the $\xi^A$ satisfy the conformal gauge constraints on $S^3$
\begin{equation}
\partial_{\tau}\xi^A\partial_{\tau}\xi^A-\partial_{\sigma}\xi^A\partial_{\sigma}\xi^A=0\,,\qquad
\partial_{\tau}\xi^A\partial_{\sigma}\xi^A=0\,.
\label{S3-conformal}
\end{equation}
In this case, $\xi^A(\tau,\sigma)$ is a minimal surface and the action (\ref{S3-action}) gives its area. 
But the Virasoro constraints for the string $\sigma$-model on $AdS_3\times S^2$, namely $T^{AdS}_{\alpha\beta}+T^{S}_{\alpha\beta}=0$, do not in general imply the vanishing of the stress tensor of the auxiliary $S^3$ $\sigma$-model.  
In fact, it is possible to show from the pseudo-holomorphicity equations that the constraints (\ref{S3-conformal}) are satisfied if and only if the Virasoro constraints on $AdS_3\times S^2$ are separately satisfied, $T^{AdS}_{\alpha\beta}=0=T^{S}_{\alpha\beta}$. This happens for example in the case of the latitude solution, and hence the corresponding solution on $S^3$ has a minimal surface interpretation, but it is clearly not the most general case.

An example of a relevant $S^3$ solution which does not have a geometric interpretation as minimal surface, is the one corresponding to the two-longitude solution. The conformal gauge constraints on $S^3$ are not met in this case, since the two-longitude solution on $AdS_3\times S^2$ only satisfies $T^{AdS}_{\alpha\beta}+T^{S}_{\alpha\beta}=0$. The explicit solution on $S^3$ is given by
\begin{equation}
\begin{aligned}
&\xi^1=-\cos a\sigma \tanh\sqrt{1-a^2}\tau\,,\qquad \xi^2=\sin a\sigma \tanh\sqrt{1-a^2}\tau\,,\\
&\xi^3=-\frac{\cos \sigma}{\cosh\sqrt{1-a^2}\tau}\,,\qquad \xi^4=\frac{\sin\sigma}{\cosh\sqrt{1-a^2}\tau}\,.
\end{aligned}
\label{S3-longitudes}
\end{equation}
Solving (\ref{x-sol2}) and (\ref{y-sol}) one can recover the corresponding supersymmetric solution on $AdS_3 \times S^2$. The constant $C$ is fixed by the constraint $x^2+z^2=1$ to be $C=\pm \frac{a}{\sqrt{1-a^2}}$. The choice of positive sign gives the solution in (\ref{longitudes}), while the choice of negative sign gives a different solution which we interpret as the unstable counterpart of (\ref{longitudes}). This unstable solution, which has not been previously discussed in the literature, is related to the stable one in (\ref{longitudes}) by $y^i \rightarrow y^i-\frac{2a}{\sqrt{1-a^2}} \frac{x^i}{x^2}$. It is easy to see that this solution has the same boundary conditions as the stable one, but wraps the ``complementary'' region on $S^2\subset S^5$ compared to (\ref{longitudes}). It can be also checked by explicit calculation that its regularized action is equal and opposite to the one of the stable solution, as expected.

To conclude this section, let us also mention that the reduction to the $S^3$ $\sigma$-model allows to rewrite the general expression for the regularized action, eq. (\ref{Sreg}), in a potentially more suggestive form. Since $z \propto \xi^4$, it follows from the $S^3$ $\sigma$-model equations of motion that 
\begin{equation}
S_{reg}=\int_{\Sigma} J_0 = \frac{1}{2}\int d^2\sigma\left(\partial_{\alpha}\theta^i \partial_{\alpha}\theta^i-\partial_{\alpha}\xi^A \partial_{\alpha}\xi^A\right)\,.
\end{equation} 
This equation is for example useful as it allows to easily repeat the deformation argument of Section \ref{deform-section} directly at the level of the regularized action. Since the equations of motion for the $\theta^i \in S^2$ and the $\xi^A \in S^3$ are separately satisfied, it follows that the variation of $S_{reg}$ is a boundary term 
\begin{equation}
\delta S_{reg}=-\frac{1}{2} \int_{\partial \Sigma}d\tau \left(\delta \theta^i \partial_{\sigma} \theta^i-\delta \xi^A \partial_{\sigma} \xi^A\right)\,.
\end{equation}
Going through the same steps as in Section \ref{deform-section}, one finally finds that $\delta S_{reg}=-C\delta A_1$, in agreement with the fact stated earlier that the regularizing boundary term does not affect the deformation calculation of Section \ref{deform-section}.

\subsection{Pohlmeyer reduction and Complex-sinh-Gordon model}

It is well known \cite{Pohlmeyer:1975nb,Lund:1976ze,Pohlmeyer:1979ch,Getmanov:1980cq} that given a $\sigma$-model on $S^n$ we can perform a reduction \`a la Pohlmeyer which reduces the dynamical variables leaving only the ``physical'' degrees of freedom. This kind of reduction was also recently considered in various string related contexts, {\it e.g.} in \cite{Chen:2006gea,Hayashi:2007bq,Grigoriev:2007bu,Mikhailov:2007xr,Jevicki:2007aa,Klose:2008rx,Grigoriev:2008jq,Miramontes:2008wt,Roiban:2009vh,Alday:2009yn}. In this section we will apply the reduction to the auxiliary sigma model on the $S^3$. This will shed a new light on some of the solutions of the pseudo-holomorphic equations. Hopefully this might also help in gaining some insight on the general properties of the solutions and to construct new ones.

Exploiting the classical conformal invariance of the $S^3$ $\sigma$-model we can always bring the components $T_{zz}$ and $T_{\bar{z}\bar{z}}$ of the stress tensor into the form
\begin{equation}\label{pohlm}
 \partial_z \xi^A \partial_z \xi^A=\partial_{\bar{z}} \xi^A \partial_{\bar{z}}\xi^A=\mu^2\,;\qquad \partial_z\doteq\partial_\sigma-i\partial_\tau\,,\partial_{\bar{z}}\doteq\partial_\sigma+i\partial_\tau\,,\end{equation}
where $\mu$ is a constant. To obtain the ``reduced'' theory we now introduce the fields $\phi$ and $\theta$ through the following non-linear field redefinition
\begin{equation}
 \begin{aligned}
 \mu^2 \cosh{2\phi} &\doteq \partial_z \xi^A \partial_{\bar{z}} \xi^A\,,\\
-\mu^3\partial_{\bar{z}}\theta \sinh^2\phi  &\doteq \frac{1}{2}\partial_{\bar{z}}^2 \xi^A K^A  \,,\\
\mu^3\partial_{z}\theta \sinh^2\phi &\doteq \frac{1}{2}\partial_{z}^2 \xi^A  K^A\,,
 \end{aligned}
\end{equation}
where the vector $K$ is defined as follows
\begin{equation}
 K^A=\epsilon_{ABCD}\xi^B \partial_{\bar{z}}\xi^C\partial_{z}\xi^D\,.
\end{equation}
The use of hyperbolic functions, compared to the trigonometric ones as in the usual Pohlmeyer reduction, is convenient in this Euclidean worldsheet setting for keeping the variable $\phi$ real. Via this procedure we have effectively eliminated one of the fields of the $S^3$ $\sigma$-model and reduced the theory to a two-dimensional one with $\theta$ and $\phi$ as dynamical variables. If we introduce the complex variable \begin{equation}
 \psi=\sinh\phi\,\exp(-i\theta/2)\,,
\end{equation}
the equations of motion for the original $\sigma$-model are satisfied provided that $\psi$ satisfies the complex-sinh-Gordon (CshG) equation
\begin{equation}\label{cshg}
 \partial_{\bar{z}}\partial_z \psi-\psi^{*}\frac{\partial_{\bar{z}}\psi\partial_z \psi}{1+|\psi|^2}+\mu^2 \psi(1+|\psi|^2)=0\,. 
\end{equation}
When $\theta$ is constant this equation reduces to the sinh-Gordon equation for the variable $\phi$
\begin{equation}\label{shg}
 \partial_{\bar{z}}\partial_z\phi+\mu^2/2 \sinh(2\phi)=0\,.
\end{equation}

After Pohlmeyer reduction, the longitude solution\footnote{We note that the latitude solution \eqref{S3-latitude} is singular from the point of view of Pohlmeyer reduction since in this case $\mu=0$.} \eqref{S3-longitudes} satisfies the CshG equation with $\mu=a$ and can be written as 
\begin{equation}
 \psi_{long}=i\,\frac{\sqrt{1-a^2}}{a}\frac{\exp(-i\sigma)}{\cosh(\sqrt{1-a^2}\tau)}\,.
\end{equation}
 Remarkably, $\psi_{long}$ is formally analogous to the soliton solution of complex-sine-Gordon theory (CsG). Indeed the solitonic solution of the latter theory can be  written as follows\footnote{The most general soliton solution can be obtained by ``boosting" eq. (\ref{soliton}).} \cite{Getmanov:1980cq,Dorey:1994mg}
\begin{equation}\label{soliton}
 \psi_{sol}=\cos\alpha\frac{\exp(i\mu \sin\alpha\, t)}{\cosh(\mu \cos\alpha\, x)}
\end{equation}
where $\alpha$ is a constant parameter. The dyonic magnon \cite{Chen:2006gea} (which is a generalization of the giant magnon solution \cite{Hofman:2006xt}) is a soliton of CsG which corresponds to $\psi_{sol}$.
The longitude solution can be recast in a form equivalent to \eqref{soliton} by taking $\mu=a$ and $\cos\alpha=\sqrt{a^2-1}/a$\footnote{This choice corresponds to $\alpha-\pi/2$ purely imaginary.}. In the case of $\psi_{long}$ the role of ``time" $t$ is taken by the variable $\sigma$ while $\tau$ is the analogue of the spatial coordinate $x$. 

The CshG theory can be thought as a generalization of sinh-Gordon with an additional internal U(1) symmetry. The CshG  solitons are charged under this symmetry.  The longitude solution $\psi_{long}$ carries  a U(1) charge which can be computed as follows
\begin{equation}
 Q=i\int d\tau \frac{\psi^*\partial_\sigma \psi-\psi\partial_\sigma \psi^*}{1+|\psi|^2}=4\, \mbox{arctanh}(\sqrt{1-a^2}).
\end{equation}
Remembering the relation $a=(\pi-\delta)/\pi$ between $a$ and the longitude angle $\delta$, we see that the charge of $\psi_{long}$ grows monotonically with $\delta$ and vanishes for $\delta\rightarrow 0$. Up to a rotation, the latitude solution corresponds to the singular limit of infinite U(1) charge.

\subsection{Unstable solutions}
\label{unstable}
In the examples of the latitude and two-longitudes loops we have explicitly seen that there are two different solutions, corresponding to the same loop at the boundary, which preserve the same supersymmetry but have equal and opposite regularized action. In this section we argue that this is a general property of all $1/8$-BPS loops on $S^2$.

Suppose we have a solution $(x^i,y^i)$ to the BPS equations (\ref{equations-explicit}), corresponding to a given supersymmetric loop on $S^2$. Then we can construct a new solution $(x^i,\tilde{y}^i)$ by keeping the same $x^i$ while changing the $y^i$ as
\begin{equation}
\tilde{y}^i = y^i-2C \frac{x^i}{x^2}\,.
\label{tilde-y}
\end{equation}
Note that the $AdS$ radial coordinate $z=\sqrt{y^{-2}}$ is left invariant by this redefinition of the $y^i$. Since $\vec{x} \times \vec{y}=\vec{x} \times \vec{\tilde{y}}$, it is clear the first line in (\ref{equations-explicit}) is still satisfied. Further, we have that $x^i \tilde y^i=-C$, so the second line of (\ref{equations-explicit}) is also satisfied, with the new integration constant simply given by $\tilde C=-C$. As argued at the beginning of Section \ref{S3-section}, the last line of (\ref{equations-explicit}) follows from the first two, so we have shown that $(x^i,\tilde{y}^i)$ is a new solution of the pseudo-holomorphicity equations. Moreover, the new solution has the same boundary conditions as the original one, since the $x^i$ are the same, while the second term in (\ref{tilde-y}) is subdominant close to the boundary, so that $z y^i|_{\sigma=0}=z \tilde{y}^i|_{\sigma=0}$. Since $C\rightarrow -C$, we expect from eq. (\ref{S-variation}) that the solutions $(x^i,y^i)$ and $(x^i,\tilde{y}^i)$ should have equal and opposite regularized areas. It would be interesting to also prove this directly by comparing the explicit expressions (\ref{Sreg}) for the two solutions. While the interpretation of this pair of solutions as stable and unstable worldsheets is natural, it would be also interesting to directly study the quadratic fluctuations around these saddle points, and show that for one of the solutions there are tachyonic modes in the three directions orthogonal to the $S^2\subset S^5$ (generalizing the observations of \cite{Drukker:2006ga} for the $1/4$-BPS circular loop). 

\section{Linearized perturbations around circular solutions}
In this section we obtain an approximate $1/8$-BPS solution by solving the pseudo-holomorphicity equations for linearized perturbations around the latitude solution. The reduction to the $S^3$ $\sigma$-model obtained in Section \ref{S3-section} turns out to be quite useful for this purpose. The results of this section give an explicit confirmation to the general argument of Section \ref{deform-section}, and they also provide an example of how to use the $S^3$ $\sigma$-model to generate $AdS_3\times S^2$ solutions.

It is convenient to take the following conformally flat metric on $S^3$
\begin{equation}
ds^2=4\frac{d\rho^2+\rho^2d\phi^2+d\zeta^2}{\left(1+\rho^2+\zeta^2\right)^2}\,.
\end{equation}
This form of the metric is related to the embedding coordinates $\xi^A$ by 
\begin{equation}
\xi^A=\left(\frac{2\rho\cos\phi}{1+\rho^2+\zeta^2},\frac{2\rho\sin\phi}{1+\rho^2+\zeta^2},
\frac{2\zeta}{1+\rho^2+\zeta^2},\frac{1-\rho^2-\zeta^2}{1+\rho^2+\zeta^2}\right)\,.
\end{equation}
The $S^3$ $\sigma$-model action in these coordinates reads
\begin{equation}
S=\frac{1}{2}\int d^2\sigma\,4\frac{\partial_{\alpha}\rho\partial_{\alpha}\rho
+\rho^2\partial_{\alpha}\phi\partial_{\alpha}\phi+\partial_{\alpha}\zeta\partial_{\alpha}\zeta}{\left(1+\rho^2+\zeta^2\right)^2}\,. 
\end{equation}
We now define the fluctuation fields around a solution $(\bar\rho,\bar\phi,\bar\zeta)$ as
\begin{equation}
\begin{aligned}
&\rho=\bar\rho+\frac{\epsilon}{2}(1+\bar\rho^2+\bar\zeta^2)\,\varphi_1\\
&\phi=\bar\phi+\frac{\epsilon}{2}\frac{1}{\bar\rho}(1+\bar\rho^2+\bar\zeta^2)\,\varphi_2\\
&\zeta=\bar\zeta+\frac{\epsilon}{2}(1+\bar\rho^2+\bar\zeta^2)\,\varphi_3\,.
\end{aligned}
\end{equation}
As it is often convenient, we have rescaled the fluctuation fields by the inverse vielbein so that $\varphi_i$ carry flat indices. We have also introduced a formal parameter $\epsilon$ to keep track of the perturbative expansion. 

To work out the linearized equations of motion for the fluctuations, one should now expand the action at quadratic order and read off the corresponding kinetic operators. Let us specialize to the case in which the solution to expand around is the one given in eq. (\ref{S3-latitude}), corresponding to the $1/4$-BPS latitude
\begin{equation}
\bar\rho=e^{-\sigma}\,,\qquad \qquad \bar\phi=\tau+\frac{\pi}{2}\,,\qquad \qquad \bar\zeta=0\,.
\label{circle-solution}
\end{equation}
Dropping total derivatives which do not contribute to the equations of motion, the action at quadratic order then turns out to be
\begin{equation}
\begin{aligned}
S_{(2)}=\frac{\epsilon^2}{2}\int d^2\sigma \Bigg{(} &(\partial_{\sigma}\varphi_1)^2+
(\partial_{\sigma}\varphi_2)^2+(\partial_{\tau}\varphi_1-\tanh\sigma\varphi_2)^2+
(\partial_{\tau}\varphi_2+\tanh\sigma\varphi_1)^2\\
&-\frac{1}{\cosh^2\sigma}(\varphi_1^2+\varphi_2^2)+\partial_{\alpha}\varphi_3\partial_{\alpha}\varphi_3
-\frac{2}{\cosh^2\sigma}\varphi_3^2\Bigg{)}\,.
\end{aligned}
\end{equation}
This can be viewed as the action for a free scalar (the transverse fluctuation $\varphi_3$)  and a free vector field (the two longitudinal fluctuations) living on the $S^2$ hemisphere with metric (\ref{half-S2}). To see this, note that the two longitudinal fluctuations can be combined into a vector field $A^a=(\varphi_1,\varphi_2)$, where $a$ is a 2d tangent space flat index. The covariant derivative with respect to the metric (\ref{half-S2}) acts as 
\begin{equation}
\nabla_{\alpha}A^a=\partial_{\alpha}A^a+\omega_{\alpha~b}^{\,\,\,a} A^b\,,
\end{equation}
where the only non-zero component of the spin connection is $\omega_{\tau~\sigma}^{\,\,\,\tau}=-\tanh\sigma$. Then the quadratic action can be written covariantly as 
\begin{equation}
\begin{aligned}
S_{(2)}=\frac{\epsilon^2}{2}\int d^2\sigma \sqrt{g}\left(g^{\alpha\beta}\nabla_{\alpha}A^a\nabla_{\beta}A^a-A^a A^a+g^{\alpha\beta}\partial_{\alpha}\varphi_3\partial_{\beta}\varphi_3-2\varphi_3^2\right)\,,
\end{aligned}
\end{equation}
or, integrating by parts and dropping total derivatives
\begin{equation}
S_{(2)}=-\frac{\epsilon^2}{2}\int d^2\sigma \sqrt{g}\left(A^a(\nabla^2+1)A^a+\varphi_3(\nabla^2+2)\varphi_3\right)\,.
\end{equation}
The equations of motion for the fluctuations
\begin{equation}
\nabla^2A^a=-A^a\,,\qquad \qquad \nabla^2\varphi_3=-2\varphi_3\,,
\label{fluct-eq}
\end{equation}
are now recognized to be the equations satisfied by vector and scalar spherical harmonics on $S^2$
\begin{equation}
\nabla^2Y_{\alpha}^{lm}=-\left[l(l+1)-1\right]Y_{\alpha}^{lm}\,,\qquad \qquad \nabla^2Y^{lm}=-l(l+1)Y^{lm}\,,
\end{equation}
with $l=1$ in both cases. The solutions are well known and can be written in terms of Legendre polynomials
\begin{equation}
Y^{lm}(\sigma,\tau)=P^m_l(\tanh\sigma)e^{im\tau}\,,\qquad Y^{lm}_{\alpha}(\sigma,\tau)
=\epsilon_{\alpha\beta}\partial^{\beta}Y^{lm}(\sigma,\tau)\,,
\end{equation}
with $|m|\le l$\footnote{There is a small subtlety here since we are looking for solutions on the $S^2$ hemisphere with $0<\sigma<\infty$ rather than the full sphere $-\infty<\sigma<\infty$. The restriction on the quantum number $m$ comes from requiring regularity on the whole sphere. If we only need regularity at $\sigma\rightarrow \infty$ and $\sigma \rightarrow 0$, then additional solutions of the form $P_l^{m}(\tanh\sigma)e^{im\tau}$ with $-\infty < m < -l$ would be allowed (here $P_l^{m}(\tanh\sigma)$ has to be thought in general as an associated Legendre function and it is not a polynomial for $m<-l$). However one can see a posteriori that including these extra modes does not lead to a solution satisfying the correct boundary conditions and the constraint $x^2+z^2=1$, so we do not include them here.}. Specializing to the case $l=1$ and taking real linear combinations, one gets the following solutions of (\ref{fluct-eq})
\begin{equation}
\varphi_3=a_0 \tanh\sigma + a_1 \frac{\cos\tau}{\cosh\sigma}+b_1 \frac{\sin\tau}{\cosh\sigma}\,,
\end{equation}
and
\begin{equation}
\begin{aligned}
&\varphi_1=A^{\tau}=\cosh\sigma\partial_{\sigma}\tilde\varphi\,,\qquad 
\varphi_2=A^{\sigma}=-\cosh\sigma\partial_{\tau}\tilde\varphi\,,\\
&\tilde\varphi=\tilde{a}_0 \tanh\sigma + \tilde{a}_1 \frac{\cos\tau}{\cosh\sigma}+\tilde{b}_1 \frac{\sin\tau}{\cosh\sigma}\,,
\end{aligned}
\end{equation}
where $a_0,\,a_1,\,b_1,\,\tilde a_0,\,\tilde a_1\,,\tilde b_1$ are arbitrary constants.

To obtain the perturbed solution in terms of the coordinates $x^i,y^i$ on $AdS_3\times S^2$, it is first convenient to go back to the embedding coordinates $\xi^A$, for which the solution to linear order in the perturbation is 
\begin{equation}
\begin{aligned}
&\xi^1=-\frac{\sin\tau}{\cosh\sigma}-\epsilon\left(\sin\tau\tanh\sigma\varphi_1+\cos\tau\varphi_2\right)\,,\\
&\xi^2=\frac{\cos\tau}{\cosh\sigma}+\epsilon\left(\cos\tau\tanh\sigma\varphi_1-\sin\tau\varphi_2\right)\,,\\
&\xi^3=\epsilon \varphi_3\,,\\
&\xi^4=\tanh\sigma -\epsilon \frac{\varphi_1}{\cosh\sigma}\,.
\end{aligned}
\end{equation}
From here one sees that the boundary condition $\xi^4 \propto z \rightarrow 0$ at $\sigma=0$ fixes $\tilde a_0=0$, while all other constants are still arbitrary. To get the solution for the $x^i$, one has then to integrate 
\begin{equation}
\partial_{\alpha}x^i =  \frac{1}{\sqrt{1+C^2}}\epsilon_{\alpha\beta}\left(\xi^4\partial_{\beta}\xi^i-\xi^i \partial_{\beta}\xi^4\right)\,.
\end{equation}
Since we are considering a perturbation of the latitude solution, the integration constant $C$ appearing here can be written in the form\footnote{Again we choose a specific sign of $C$ corresponding to the stable latitude solution. One can analogously obtain a perturbed unstable latitude solution.}
\begin{equation}
C=\cot\theta_0+\epsilon \gamma + {\mathcal O}(\epsilon^2)\,,
\end{equation}
for some constant $\gamma$. One can now solve the above differential equation to linear order in $\epsilon$ and fix the resulting integration constants so that the constraint $x^2+z^2=1$ and the consistency condition $\xi^i x^i=0$ are satisfied. The latter constraint fixes $a_0=0$, and one obtains the perturbed solution for the $x^i$ as a function of the free parameters $\gamma,\,a_1,\,b_1,\,\tilde a_1,\,\tilde b_1$ and the latitude angle $\theta_0$. The explicit expression is a bit lengthy and we do not write it out here. The loop at the boundary $x^i_{(0)}=x^i(\tau,0)$ is some kind of ``wavy" latitude whose area can be computed to be
\begin{equation}
A_1 = \int_0^{2\pi} d\tau \frac{x^1_{(0)} \dot x^2_{(0)}-x^2_{(0)}\dot x^1_{(0)}}{1+x^3_{(0)}}=2\pi\left(1-\cos\theta_0-\epsilon \gamma \sin^3\theta_0\right)+{\mathcal O}(\epsilon^2)\,.
\label{area-epsilon}
\end{equation}
Note that the area only depends on the constant $C$ consistently with the general relation (\ref{CC}), though the shape of the loop and of the worldsheet clearly depends on all other constants.

Having determined the $\xi^A$ and the $x^i$, the $y^i$ are explicitly obtained as explained above from eq. (\ref{y-sol}). The resulting solution $(x^i,y^i)$ satisfies the pseudo-holomorphicity equations up to quadratic order in $\epsilon$.

The regularized area of the solution can be computed from (\ref{Sreg}) and the result 
to linear order in $\epsilon$ turns out to be 
\begin{equation}
S_{reg}=-\sqrt{\lambda}\sin\theta_0\left(1-\epsilon \gamma \sin\theta_0 \cos\theta_0\right)+{\mathcal O}(\epsilon^2)\,. 
\end{equation}
Comparing with (\ref{area-epsilon}), one finds agreement with the general formula $S_{reg}=-\frac{\sqrt{\lambda}}{2\pi}\sqrt{A_1A_2}$.
 
\section{SUSY solutions for coincident latitudes}
\label{coincident-lat}
The reduction of the holomorphicity conditions to the $S^3$ $\sigma$-model can be used to obtain a new supersymmetric solution, which we interpret as corresponding to the correlator of two coincident latitudes with opposite orientations. This can be obtained by adapting to our case a solution on $S^3$ given in \cite{Dymarsky:2006ve}.

Start with the following parametrization of $S^3$
\begin{equation}
\xi^A=\left(\cos\eta \cos\phi,\cos\eta \sin\phi,\sin\eta\cos\psi,\sin\eta\sin\psi\right)\,,
\end{equation}
and take the circular symmetric ansatz (the shift by $\pi/2$ in $\phi$ is for later convenience)
\begin{equation}
\phi=\tau+\frac{\pi}{2}\,,\qquad \qquad \eta=\eta(\sigma)\,,\qquad \psi=\psi(\sigma)\,.
\label{coinc-ansatz}
\end{equation}
The $\sigma$-model action reduced on this ansatz reads
\begin{equation}
S=\frac{1}{2}\int d^2\sigma\left(\eta '^2+\sin^2\eta\psi '^2 +\cos^2\eta\right)\,.
\end{equation} 
There are two conserved quantities
\begin{equation}
\begin{aligned}
&p_{\psi}=\sin^2\eta \psi ' \\
&E=\frac{1}{2}\left(\eta '^2+\sin^2\eta\psi '^2 -\cos^2\eta\right)\,.
\end{aligned}
\end{equation}
We now restrict to the case in which $\psi '=0$ and take $\psi=\pi/2$. Then $\eta(\sigma)$ is determined by
\begin{equation}
\eta '^2=\epsilon^2-\sin^2\eta\,,
\end{equation}
where we have defined $\epsilon^2=2E+1>0$. If $\epsilon=1$ this differential equation is elementary and it is solved by $\sin\eta=\tanh\sigma$, which is the solution (\ref{S3-latitude}), and thus corresponds to the single latitude solution. For general $\epsilon$ the above differential equation can be solved in terms of elliptic functions
\begin{equation}
\sigma=\pm \int_0^{\eta(\sigma)} \frac{d\eta}{\sqrt{\epsilon^2-\sin^2\eta}}=\pm \frac{1}{\epsilon} F\left(\eta(\sigma),\frac{1}{\epsilon^2}\right)\,.
\label{elliptic}
\end{equation}
The behavior of the solution is different depending on whether $\epsilon>1$ or $\epsilon<1$ and the undetermined $\pm$ sign must be chosen accordingly. 

For $\epsilon>1$, the derivative $\eta '$ never vanishes, and $\eta(\sigma)$ increases monotonically (corresponding to the positive sign in (\ref{elliptic})) from $\eta(0)=0$ to $\eta(\sigma_f)=\pi$, where $\sigma_f=\frac{1}{\epsilon} F\left(\pi,\frac{1}{\epsilon^2}\right)$. Correspondingly $\sin\eta$ (and hence the $AdS$ coordinate $z$), starts at zero at $\sigma=0$, reaches a maximum $\sin\eta=1$ at $\sigma=\frac{\sigma_f}{2}$ and then vanishes again at $\sigma=\sigma_f$. The topology of the solution is thus that of a cylinder, since there are two boundaries.  

For $\epsilon<1$ the derivative $\eta '$ vanishes at $\eta=\eta_{\mbox{\tiny MAX}}=\arcsin\epsilon$, therefore one chooses the positive sign in (\ref{elliptic}) for $0<\sigma<\frac{\sigma_f}{2}$ and the negative one for $\frac{\sigma_f}{2}<\sigma<\sigma_f$, where now the final value of $\sigma$ is given by $\sigma_f = \frac{2}{\epsilon}F\left(\arcsin{\epsilon},\frac{1}{\epsilon^2}\right)$. The coordinate $\eta$ therefore starts at zero, reaches a maximum $\eta_{\mbox{\tiny MAX}}<\frac{\pi}{2}$ and then turns back to reach again $\eta=0$. Also in this case $\sin\eta$, and hence $z$, has two zeroes, so the topology is again that of a cylinder.

Through equations (\ref{x-sol2}) and (\ref{y-sol}) we can now reconstruct the full supersymmetric solution in $AdS_3\times S^2$. Firstly it is clear that 
\begin{equation}
z=\frac{\sin\eta}{\sqrt{1+C^2}}\,.
\end{equation}
Integrating (\ref{x-sol2}) one gets 
\begin{equation}
\begin{aligned}
x^1=s(\epsilon)\frac{\sqrt{\epsilon^2-\sin^2\eta}}{\sqrt{1+C^2}} \cos\tau\,,\qquad x^2=s(\epsilon)\frac{\sqrt{\epsilon^2-\sin^2\eta}}{\sqrt{1+C^2}} \sin\tau\,,\qquad x^3=\frac{\sqrt{1+C^2-\epsilon^2}}{\sqrt{1+C^2}}\,,
\end{aligned}
\end{equation} 
where $s(\epsilon)=1$ for $\epsilon \ge 1$ and $s(\epsilon)=\mbox{sign}(\frac{\sigma_f}{2}-\sigma)$ for $\epsilon<1$. For all values of $\epsilon$, the curve at both boundaries of the cylinder is the same latitude circle with angle $\theta_0$ related to $C$ by\footnote{For simplicity we pick here the positive sign of $C$. There is an analogous solution corresponding to the choice of negative $C$.}
\begin{equation}
C=\frac{\sqrt{\epsilon^2-\sin^2\theta_0}}{\sin\theta_0}\,.
\end{equation}
Note that for $\epsilon<1$ the solution exist only as long as $\epsilon>\sin\theta_0$.

The $y^i$ coordinates can be obtained from 
\begin{equation}
\vec{y}=\frac{1+C^2}{C^2+\cos^2\eta}\left(\frac{\sqrt{1+C^2}}{\sin\eta} \vec{x}\times\vec{\xi}+C \vec{x}\right)\,,
\end{equation}
where $\vec{\xi}=(\xi^1,\xi^2,\xi^3)$. The explicit expression is a bit lengthy and we do not write it out here. It is possible to explicitly verify that the full solution $(x^i,y^i)$ correctly satisfies all the pseudo-holomorphicity equations (\ref{equations-explicit}). In both the $\epsilon<1$ and $\epsilon>1$ cases it can be seen that the loop ends on opposite latitude circles on the $S^2 \subset S^5$, namely on $zy^3=\sin\theta_0$ at $\sigma=0$ and on $zy^3=-\sin\theta_0$ at $\sigma=\sigma_f$. This difference in the scalar coupling is consistent with the supersymmetric boundary conditions and is due to the opposite orientations of the two coincident latitudes in space-time.  

Though in both cases the solution describes a correlator between coincident latitudes, the behavior of the solution in the bulk is qualitatively different for $\epsilon>1$ and $\epsilon<1$. In particular, for $\epsilon <1$ the $(x^1,x^2)$ circle shrinks to zero size in the bulk while the $(y^1,y^2)$ never does, and the opposite happens in the $\epsilon>1$ case. In the limiting case $\epsilon=1$ both circles shrink to zero size at the midpoint.

It is not difficult to evaluate the regularized area of this solutions by using eq. (\ref{Sreg}). 
Plugging in the solution and evaluating the integral one finds that, both for $\epsilon>1$ and $\epsilon<1$, the contribution from the region $0\le\sigma\le\frac{\sigma_f}{2}$ is exactly cancelled by the contribution coming from the other half $\frac{\sigma_f}{2}\le\sigma\le\sigma_f$. So the conclusion is that in both cases 
\begin{equation}
S_{reg}=0\,.
\label{saddle-connected}
\end{equation}
Note that the value of the action, as well as the boundary conditions, do not depend on the parameter $\epsilon$. We interpret this parameter as a modulus which should be integrated over with the appropriate measure in the string partition function, but we do not investigate this further here.

Besides the connected solution described above, one can also consider ``degenerate'' solutions for correlators of coincident latitudes using the known 1/4-BPS latitude (\ref{latitude}). A solution with the topology of the cylinder can be obtained by joining two disks of the type (\ref{latitude}), one for each of the two coincident loops, by a thin zero-area tube. Physically, this should be interpreted as the exchange of light supergravity modes between the two worldsheets (note that the two worldsheets are not coincident because the scalar coupling is different due to the opposite orientation). The area of the solution is just going to be equal to the sum of the two areas, while the amplitude for the supergravity mode exchange will contribute as a prefactor of the exponential in the string partition function. By coupling this way the stable/unstable versions of the solution (\ref{latitude}), one obtains worldsheets with regularized area
\begin{equation}
S_{reg}=-2\sqrt{\lambda}\sin\theta_0\,,\qquad S_{reg}=2\sqrt{\lambda}\sin\theta_0\,,\qquad
S_{reg}=0\,.
\label{saddles-coinc}
\end{equation}
The solution obtained by gluing the stable and unstable worldsheet can be thought as the limiting case $\epsilon\rightarrow 1$ of the above family of connected solutions with zero regularized area. The first choice with $S_{reg}=-2\sqrt{\lambda}\sin\theta_0$ is clearly the dominant contribution at strong coupling. However we will see that all three possibilities appear as saddle points at strong coupling in a matrix model computation. 

\subsection{The coincident correlator as soliton of sin(h)-Gordon}\label{pohlmeyersec}

Also for the solution (\ref{coinc-ansatz})-(\ref{elliptic}) contributing to the coincident latitude correlator we can apply the Pohlmeyer reduction and interpret it as a solution of CshG. In the present case it reads
\begin{equation}
 \psi_{cor.}=\frac{\mbox{cn}(\epsilon \sigma,\epsilon^{-2})}{\sqrt{\epsilon^2-1}}\exp{\left(-i\theta_0/2\right)}\,,
\end{equation}
where $\mbox{cn}(\epsilon \sigma,\epsilon^{-2})$ denotes a Jacobi elliptic function. Since $\theta=\theta_0$ is constant,  this solution has zero U(1) Noether charge. Given a chargeless solution $\psi=\sinh\phi\,\exp(-i\theta_0/2)$ of CshG, $\phi$ is automatically a solution of sinh-Gordon and in the case of the correlator we have\footnote{For $\epsilon<1$ it is convenient to add $ i\pi/2$ to \eqref{kinktrain} to keep $\phi$ real.}
\begin{equation}\label{kinktrain}
\phi=\mbox{arcsinh}\left(\frac{\mbox{cn}(\epsilon\sigma,\epsilon^{-2})}{\sqrt{\epsilon^2-1}}\right)
\end{equation}
which is indeed a solution of eq. \eqref{shg} with $\mu^2=\epsilon^2-1$. 
This expression is formally analogous, upon analytic continuation, to a ``kink-wave" periodic solution of the sine-Gordon equation (see {\it e.g.} \cite{Hayashi:2007bq,Bakas:2002qi}) which can be written, in the static limit, as
\begin{equation}
 \arcsin(\mbox{cn}(k x,\frac{1}{k^2}))\,.
\end{equation}
The connection with sine-Gordon solitons can be seen directly, even without performing Pohlmeyer reduction, by noticing that the variable $\eta$ appearing in the circular symmetric ansatz satisfies the sine-Gordon equation $\partial^2_\sigma\eta+\sin\eta\,\cos\eta=0$. The solutions relevant for the correlator are obtained by inverting \eqref{elliptic} and are of the form\footnote{For $\epsilon<1$ one has to invert the sign of $\eta$ for $\sigma_f/2<\sigma<\sigma_f$.}
\begin{equation}
 \eta=\mbox{am}(\epsilon \sigma,1/\epsilon^2)\,,
\end{equation}
which is another form of the kink-wave solution present in the literature, see e.g. \cite{Klose:2008rx}.
Using $\eta$ as variable allows to safely perform the single latitude limit $\epsilon\rightarrow 1$ to obtain
\begin{equation}
 \eta=\arcsin(\tanh(\sigma))
\end{equation}
which is the familiar expression for the kink solution of sine-Gordon.

\section{Correlators of non-coincident loops}
\label{correl-non-coinc}
It is natural to ask whether there are supersymmetric string solutions describing connected correlators between two generic loops. In this section we will provide evidence that a connected smooth configuration does not exist, the only possible surface being a degenerate one, obtained by joining the worldsheets of the individual loops by an infinitesimal tube corresponding to an exchange of supergravity modes\footnote{In the case of two 1/4-BPS latitudes, the same conclusion was previously reached independently by Nadav Drukker \cite{Nadav:private}.}. We will first start by showing that a connected worldsheet does not exist for the case of the correlator of two latitudes loops. Using a ``deformation" argument we will then argue that supersymmetric connected solutions do not exist for loops of arbitrary shape.

Let us begin by considering the connected correlator of two 1/4-BPS latitudes with latitude angles, say, $\theta_1$ and $\theta_2$. To find the associated string we should look for a solution to the pseudo-holomorphicity equations with the topology of a cylinder whose boundaries are given by the two latitude circles. Because of circular symmetry, we can take a periodic ansatz on $AdS_3\times S^2$ of the type considered in \cite{Drukker:2005cu} 
\begin{equation}
\begin{aligned}
&x^i=\tanh\rho\left(\sin u \cos\psi,\sin u \sin\psi,\cos u\right)\,,\qquad z=\frac{1}{\cosh\rho}\,,\\
&y^i=\cosh\rho\left(\sin v\cos\phi,\sin v\sin\phi,\cos v\right)\,,
\end{aligned}
\end{equation}
with\footnote{One may in general take arbitrary winding numbers $\psi=k_1\tau+\alpha(\sigma)\,,\, \phi=k_2\tau+\beta(\sigma)$, but the supersymmetric boundary conditions require $k_2=k_1$, and we take the single wrapping case $k_1=1$.}
\begin{equation}
\begin{aligned}
&\rho=\rho(\sigma)\,,\qquad u=u(\sigma)\,,\qquad v=v(\sigma)\,,\\
&\psi=\tau+a(\sigma)\,,\qquad \phi=\tau+b(\sigma)\,.
\end{aligned}
\end{equation}
The relevant solution should have $x^3(\sigma)=\tanh\rho\cos u$ varying continuously from $x^3(0)=\cos\theta_1$ to $x^3(\sigma_f)=\cos\theta_2$. However it is not difficult to see that the pseudo-holomorphicity equations do not allow such a solution. If we take the first line of eq. (\ref{equations-explicit}) with $\alpha=\tau$ and plug in the above ansatz with circular symmetry we obtain
\begin{equation}
\partial_{\sigma}x^3=0\,.
\end{equation}
So $x^3$ has to be constant and we fall back to the single latitude solution (or the coincident latitudes of the previous section). We then conclude that there is no connected supersymmetric solution describing the correlator of non-coincident latitudes. The only relevant supersymmetric worldsheets will then be degenerate cylinders constructed by joining two standard latitude solutions (\ref{latitude}) by a zero area tube, as described at the end of Section \ref{coincident-lat}, corresponding to the propagator for the exchange of supergravity modes between the two worldsheets. To leading order in the $\alpha '$ expansion, the connected correlator will be given by the exponential of the regularized area of this degenerate surface, which is just the sum of the areas of the two disks. Considering as before all possibilities of stable/unstable pairings, we then find four different saddle points with regularized areas
\begin{equation}
S_{reg}=\sqrt{\lambda}(\pm\sin\theta_1\pm\sin\theta_2)\,.
\label{4-saddles}
\end{equation}
We will show in Section \ref{2MM-section} that a Gaussian two-matrix model calculation precisely exhibits in the strong coupling limit the presence of these four saddle points. 

We now discuss the general case of a correlator between two generic distinct loops of areas $A_1$ and $A_2$. Let us {\it suppose} that a connected surface with cylinder topology joining these two loops does exist. We can then act with an area preserving deformation on the loops to bring ourselves to the case where we have two latitudes whose angles $\theta_i$ are determined by $A_1$ and $A_2$. As a consequence, the initial connected surface will be deformed into the one describing a latitude-latitude correlator. As proved before, this final surface must correspond to a disconnected worldsheet. This therefore implies that, while deforming the loops, we have moved from an initial connected worldsheet to a disconnected one.
This fact signals the presence of a discontinuity, i.e. a phase transition \cite{Gross:1998gk,Zarembo:1999bu,Olesen:2000ji}, at some intermediate configuration as we interpolate from the initial loops to the latitude ones. This phase transition, which is typically of first order, should manifest itself as a singularity of the free energy of the string. To leading order in the supergravity approximation, the string free energy is given by the worldsheet area. But by assumption we kept the area of the loops constant during the deformation and, for our class of loops, by the arguments of Section \ref{deform-section} this in turn implies that the worldsheet area of the string representing the correlator is also constant as we deform the loops. As a consequence the free energy cannot change and therefore no singularities can possibly appear during the deformation. This leads to conclude that our initial assumption was inconsistent and that as a consequence the connected surface joining two loops with the topology of two latitudes does not exist. Hence, we conclude that the connected correlator is described at strong coupling by the four saddle points (\ref{4-saddles}), with $\sin\theta_i\rightarrow \sqrt{\frac{4 A_i(A-A_i)}{A^2}}$ (the dominant contribution being, of course, the one corresponding to the choice of two negative signs).

\section{Matrix model computation}
\label{Matrix-section}

In this section we study connected correlators 
of Wilson loops in the zero instanton sector of the two-dimensional Yang-Mills on a two-sphere. 
Two-dimensional Yang-Mills theory is a well-known and explicitly solvable theory~\cite{Migdal:1975zf, Kazakov:1980zi, Kazakov:1981sb, Rusakov:1990rs, Blau:1993hj,Blau:1991mp, Gross:1989fc,Fine:1990zz,Witten:1991we,Gorsky:1993pe,Gross:1994ub}.

The exact partition function of the 2d Yang-Mills on arbitrary Riemann surface $\Sigma$ can be obtained
by gluing the basic building blocks of three types: 
the partition function of 2d Yang-Mills on a disk, on a cylinder, and on a sphere with three holes. 
Each of these partition functions depends on the boundary conditions of the gauge field, and in the gluing process one integrates over the boundary conditions. For a boundary with topology of a circle $S^1$ the boundary conditions for the gauge field modulo gauge transformations are parameterized by the holonomy of the gauge field around $S^1$. The holonomy takes values in the gauge group $G$. Hence, if $\Sigma$ is glued from the basic building blocks along $K$ circles we get the partition function of the 2d Yang-Mills on $\Sigma$ represented by an integral over $K$ copies of the group manifold $G$. Of course, one can insert Wilson loop observable on any $k$-th gluing circle, which will be simply represented by inserting a character of the corresponding $k$-th integration variable in the integral over $K$ copies of $G$. 
We will refer to the partition function of 2d Yang-Mills represented in this form as the unitary $K$-matrix model. 

In the simplest case of a sphere with possible insertion of one non-self-intersecting Wilson loop we get 
a unitary 1-matrix model. A Lie algebra $\g$ of Lie group $G$ can be viewed as a perturbative approximation to $G$.
For $G = \U(N)$ the Lie algebra $\g$ is represented by Hermitian matrices multiplied by $\sqrt{-1}$.

The connection between the zero-instanton sector of two-dimensional Yang-Mills on $S^2$ 
and the Hermitian matrix model was shown in \cite{Bassetto:2001mf,Bassetto:1999dg,Bassetto:1998sr,Staudacher:1997kn}.
Intuitively, discarding instantons in 2d Yang-Mills exactly corresponds to discarding the non-trivial topology of $G$,
and replacing $G$ by its tangent space at the unity, i.e. by the Lie algebra of $G$. 
The exact partition function of 2d Yang-Mills on a disk, which is a heat kernel for the standard Laplacian on $G$, is replaced by the heat kernel of the standard Laplacian on $\g$. But the heat kernel on $\g$ is a simple Gaussian function. Therefore one relates the zero-instanton sector of 2d Yang-Mills on a sphere with one non-self-intersecting Wilson loop to the Hermitian matrix model. For concentric Wilson loops on a sphere the argument goes in the same way, and we explicitly derive the corresponding Hermitian multi-matrix model in Section \ref{2MM-section}. 

\subsection{Conventions}

We take Euclidean space-time $\BR^4$ and assume that the gauge theory action is normalized as 
\begin{equation}
  \label{eq:4d-action}
  S_{YM} = - \frac {1} {g_{YM}^2} \int d^4 x \tr \lb \frac 1 2 F_{\mu \nu}F^{\mu \nu} + D_{\mu} \Phi_{i} D^{\mu} \Phi^i 
+ \frac 1 2 [\Phi_i, \Phi_j][\Phi^i, \Phi^j] + \dots\rb
\end{equation}
where the symbol `$\tr$' is a negatively defined invariant bilinear form on $\g$. For a simple group $G$ the choice of 
the bilinear form is unique up to a scaling. For the gauge group $G=\U(N)$ we take `$\tr$' to be the trace in the fundamental representation, 
hence the notation. Assuming that $\{T^{a}\}, a = 1 \dots \dim \g$ is the orthogonal basis on $\g$, 
and $A^a$ are the real coordinates on $\g$ in this basis, i.e. $A = A^a T_{a}$, 
and assuming the normalization $ \tr T^a T^b = -\frac 1 2 \delta_{a b}$ the propagator in the Feynman 
gauge is 
\begin{equation}
  \label{eq:propagator}
  \langle A^{a}_{\mu} (x) A^{b}_{\nu}(y) \rangle = \frac {g_{YM}^2} {4 \pi^2} \frac {\delta_{ab} g_{\mu \nu} } {( x - y )^2}.
\end{equation}

Our conventions are such that the gauge field $A$ takes value in the Lie algebra of the gauge theory, e.g. 
in the anti-Hermitian $N\times N$ matrices if the gauge group is $U(N)$. The covariant derivative is $D = d + A$ 
and the gauge field strength is the two-form $F = dA + A \wedge A$. 

For a $G$-gauge bundle over the space-time $M$ and two points $x_1,x_2 \in M$ connected by a path ${\cal C}$, the
holonomy of a gauge connection $A$ is  a parallel transport of the gauge bundle fiber from the point $x_2$
to the point $x_1$\footnote{The path ordering is defined in  such a way that if $x_{m}$ is any point on ${\cal C}$ 
then $G(x_1,x_2) = G(x_1,x_m)G(x_m,x_2)$.}
\begin{equation}
  \label{eq:parallel-transport}
  G(x_1,x_2,{\cal C})  = \Pexp \int_{x_1}^{x_2} A_{\mu} dx^{\mu}\,.
\end{equation}
A holonomy $G(x_1,x_2,{\cal C})$ takes values in the gauge group $G$. For a closed contour ${\cal C}$ and a representation $R$ of the gauge 
group the Wilson loop operator is the $R$-character of the holonomy around ${\cal C}$
\begin{equation}
  \label{eq:wilson-loop}
  W_{R}({\cal C}) = \tr_R \Pexp \oint_{{\cal C}} A_{\mu} dx^{\mu}.
\end{equation}

In ${\cal N}=4$ SYM it is natural to generalize the definition of Wilson loop operator by adding couplings to the scalar fields of the ${\cal N}=4$ multiplet. This allows to define operators which globally preserve some fraction of the superconformal symmetries. In particular, here we consider the $1/8$-BPS Wilson loops for contours ${\cal C}$ restricted to a unit sphere $S^2 \subset \BR^4$, which are obtained as a special case of the general loops in (\ref{S3-loops}) by restricting to a great two-sphere inside $S^3$. They couple to the three scalar fields as
\begin{equation}
  \label{eq:define-loop}
  W_{R}({\cal C}) = \tr_{R} \Pexp \oint_{{\cal C}} (A_{j}  + i \ep_{ijk} \Phi^{i} x^{k}) dx^{j}\,.
\end{equation}
These Wilson loops can be thought of as holonomies of the complexified connection 
 $ \tilde A_{c} = A + i * \Phi $ (here $*$ is a two-dimensional Hodge star on $S^2$) in the 
2d Higgs-Yang-Mills theory \cite{Pestun:2008thesis,Pestun:2009nn}
\begin{equation}
  \label{eq:complexified-connection}
  W_{R}({\cal C}) = \tr_{R} \Pexp \oint_{{\cal C}} \tilde A_{c}\,. 
\end{equation}

In \cite{Drukker:2007qr,Drukker:2007yx,Drukker:2007dw}, it was conjectured that the expectation value of a non-self-intersecting Wilson loop (\ref{eq:define-loop}) can be computed in the Gaussian matrix model with coupling constant which depends on the area on $S^2$ enclosed by the loop
\begin{equation}
  \label{eq:matrix-model-conjecture}
  \langle W_{R}({\cal C}) \rangle = \frac 1  Z \int_{I\g} [d \phi] 
e^{ - \frac {2}{\tilde g_{MM}^2 } \tr \phi^2} \tr_{R} e^{\phi},
\end{equation}
where the partition function $Z$ is, of course, fixed by saying that $\langle W_{R}({\cal C}) \rangle = 1$ if $R$ is the trivial 
representation.  In (\ref{eq:matrix-model-conjecture}) the integral is taken 
over the imaginary slice of the complexified Lie algebra $\g_{\BC}$, i.e. over Hermitian matrices in the $\U(N)$ case. The $\tilde g_{MM}^2$ is the area-dependent effective coupling constant in the matrix model 
\begin{equation}
  \tilde g_{MM}^2 = \frac {4 A_1 A_2} {A^2} g_{YM}^2.
\end{equation}
By $A= 4 \pi r^2$ we denote the total area of $S^2$, and $A_1, A_2$ are the 
areas inside and outside the loop ${\cal C}$ on $S^2$; they satisfy $A_1 + A_2 = A$. 
For example, specializing to the case of the latitude at the  polar  angle $\theta$ measured from the North pole we have \cite{Drukker:2006ga,Drukker:2006zk}
\begin{equation}
  \label{eq:latitude}
  \tilde g_{YM}^2(\theta) = g_{YM}^2  \sin^2 {\theta},
\end{equation}
so $\theta = \pi/2$ corresponds to the equator and gives the standard $\frac 1 2$-BPS Wilson 
loop \cite{Erickson:2000af,Drukker:2000rr}.

To streamline notations for the matrix model computation we denote   
\begin{equation}
\label{eq:matrix-model-coupling}
  \frac 2 {\tilde g_{MM}^2}  :=  \frac {N}{2 \nu}; \quad   \tilde \lambda =\tilde g_{MM}^2 N = 4 \nu\,, 
\end{equation}
then the matrix model has the form 
\begin{equation}
  \label{eq:matrix-model-nu}
\langle W_{R}({\cal C}) \rangle = \frac {1} {Z}  \int_{I \g} [ d \phi]  e^{ -\frac N {2 \nu}  \tr \phi^2 } \tr_{R} e^{\phi}
\end{equation}
so $\nu$ has a meaning of effective 't Hooft coupling constant in the matrix model.

The arguments which lead to the matrix model conjecture are exactly the same if we consider
several Wilson loop on top of each other in arbitrary representations, hence
we have
\begin{equation}
  \label{eq:matrix-model-nu-many}
\langle W_{R_1}({\cal C}) W_{R_2} ({\cal C}) \dots W_{R_k}({\cal C}) \rangle = \frac {1} {Z}  \int_{I\g} [ d \phi]  e^{ -\frac N {2 \nu}  \tr \phi^2 } \tr_{R_1} e^{\phi} \tr_{R_2} e^{\phi} \dots \tr_{R_k} e^{\phi}\,.
\end{equation}


\subsection{The connected correlator of two coincident Wilson loops}

\newcommand{\conn}{{\mathrm {conn}}}

We are going to compute the connected correlator of two Wilson loops on the same base contour ${\cal C}$
\begin{equation}
  \label{eq:connected-correlator}
\langle W_{R_1}  W_{R_2} \rangle_\conn = \langle W_{R_1} W_{R_2}  \rangle - \langle W_{R_1} \rangle \langle W_{R_2} \rangle\,.
\end{equation}
We specialize to the case where $R_1=f$ is the fundamental representation of $G=\U(N)$, 
and $R_2=\bar f$ is the dual (complex-conjugate) representation to $R_1$. This corresponds to two coincident loops of opposite orientations\footnote{The case of same orientation corresponds to $R_2=f$ and it is also easily obtained from the following calculation.}. The large $N$ expansion of the two-point correlator in the Hermitian matrix model has been previously considered in \cite{Akemann:2001st}. For results at finite $N$, see \cite{Kawamoto:2008gp}.

The $k$-point resolvent function is defined as 
\begin{equation}
  \label{eq:resolvent}
  G(z_1, \dots, z_k) = N^{k-2} \langle \tr \frac {1}{z_1 - \phi} \dots \tr \frac{1}{z_k - \phi} \rangle_\conn\,.
\end{equation}
This function is a generating function for all connected correlators of the form 
\begin{equation}
  \label{eq:connected-correlators-traces}
  \langle \tr \phi^{i_1} \tr \phi^{i_2} \dots \tr \phi^{i_k} \rangle_\conn,
\end{equation}
since
\begin{equation}
  \label{eq:resolvent-explicit}
  G(z_1, \dots, z_k) = N^{k-2} \sum_{i_1=0}^{\infty} \dots \sum_{i_k=0}^{\infty} z_1^{-1 - i_1} \dots z_k^{-1 - i_k}
  \langle \tr \phi^{i_1} \tr \phi^{i_2} \dots \tr \phi^{i_k} \rangle_\conn\,.
\end{equation}
Another generating function $W(x_1, \dots, x_k)$ is 
\begin{equation}
  \label{eq:W-gen-function}
  \begin{split}
  W(x_1, \dots, x_k) = N^{k-2} \langle \tr e^{x_1 \phi} \dots \tr e^{x_k \phi} \rangle_\conn = \\
   = N^{k-2} \sum_{i_1=0}^{\infty} \dots \sum_{i_k=0}^{\infty} x_1^{i_1}  \dots x_k^{i_k}
  \langle \frac {1} {i_1! \dots i_k!} \tr \phi^{i_1} \tr \phi^{i_2} \dots \tr \phi^{i_k} \rangle_\conn\,.
  \end{split}
\end{equation}

The function $W(x_1, \dots, x_k)$ can be obtained from $G(z_1, \dots, z_k)$ by the inverse Laplace  transform
in all variables $z_1, \dots, z_k$ 
\begin{equation}
  \label{eq:inverse-Laplace-many}
  W(x_1, \dots, x_k) = \frac {1} { (2 \pi i)^k} \oint dz_{1} \dots \oint dz_k G(z_1, \dots, z_k) e^{x_1 z_1 + \dots x_k z_k}\,,
\end{equation}
where the integration contours encircle the point $z= \infty$ for all integration variables.

The $k$-point resolvent function can be expanded in inverse powers of $N$,
which is the same as the topological expansion over the genus $h$ of Riemann surfaces associated with Feynman ribbon graph diagrams
\begin{equation}
  \label{eq:genus-expansion}
  G(z_1, \dots, z_k) = \sum_{h=0}^{\infty} G_h(z_1, \dots, z_k) N^{-2h}\,.
\end{equation}
The leading order term $h=0$ corresponds to the planar diagrams.

The genus zero $k$-point resolvent can be computed recursively from the loop equations \cite{Ambjorn:1992gw,Alexandrov:2003pj,Ambjorn:1990ji}.
The $1$-point and $2$-point genus zero correlation function are 
\begin{equation}
  \label{eq:2-point}
\begin{aligned}
  G_0(z_1) &= \frac 1 {2\nu} (z_1 - \sqrt{z^2 - 4 \nu})\,, \\
  G_0(z_1, z_2) &= \frac {1} {2 (z_1 - z_2)^2} \left ( \frac {z_1 z_2 - 4 \nu} { \sqrt{z_1^2 - 4\nu} \sqrt{z_2^2 - 4\nu}} - 1 \right)\,.
\end{aligned}
\end{equation}

To compute Wilson loop correlators we need to perform the inverse Laplace transform. For the $1$-point function the
result is 
\begin{equation}
  \label{eq:inverse-laplace-single}
  W_0(x) = \frac {I_1(2 \sqrt {\nu} x)} {\sqrt {\nu} x} = 1+\frac{\nu  x^2}{2}+\frac{\nu ^2 x^4}{12}+\frac{\nu ^3 x^6}{144}+ \dots 
\end{equation}
where $I_1$ is the modified Bessel function of the first kind.
For a single Wilson loop in the fundamental representation we take $x = 1$ and recover the well-known result \cite{Erickson:2000af}
\begin{equation}
  \label{eq:single-loop}
  \langle \frac 1  N W_f({\cal C}) \rangle  = \frac {I_1 (2 \sqrt{\nu})}{ \sqrt{\nu} }  = 
\frac {2 I_{1} (\sqrt{ \tilde \lambda})} { \sqrt {\tilde \lambda}}\,.
\end{equation}

The large $\nu$ asymptotic can be extracted from the integral representation for the modified Bessel function $I_n(x)$\footnote{
The modified Bessel function $I_n$ is related to the Bessel function $J_n$ of imaginary argument $I_n(x) = (-i)^n J_n(ix)$.}
\begin{equation}
  \label{eq:integral-repre}
  I_n(x)  = \frac 1 { 2\pi} \int_{-\pi}^{\pi} e^{x \cos (\theta )} \cos (n \theta ) \, d\theta\,.
\end{equation}
At large $x$ we can use the semi-classical approximation. There are two critical points $\theta=0$ and $\theta = \pi$.
Expanding in $\theta$ near $\theta = 0$ and taking the integral perturbatively we get
\begin{equation}
  \label{eq:expansion0}
  I_n(x) = \frac {1} {\sqrt{ 2 \pi x} } e^{x} \left( 1+\frac{1-4 n^2}{8 x}+\frac{1}{128 x^2} \left(16 n^4-40 n^2+9\right) 
+ \dots \right)\,.
\end{equation}
Expanding in $\theta$ near $\theta = \pi$ we get get
\begin{equation}
\label{eq:expansionPi}
I_n(x) = \frac { i  e^{-x+ n \pi i } } {\sqrt{ 2 \pi x}} \left(1 - \frac{1-4n^2}{8x}+ \frac{1}{128 x^2} \left(16 n^4-40 n^2+9\right) + \dots \right)\,.
\end{equation}
Notice the (\ref{eq:expansionPi}) is imaginary for real $x$ because we make the expansion around the unstable critical point\footnote{The full asymptotic series is \cite{Andrews:1999:SF} p. 223
\[
I_n(x) = \frac {e^x} {\sqrt{2\pi x}} \sum_{k=0}^{\infty} \frac {(-1)^{k} (n,k)}{ (2x)^k} + \
\frac {e^{-x \pm (n + \frac 1 2 ) \pi i}}{ \sqrt{ 2 \pi x}} \sum_{k=0}^{\infty} \frac { (n,k)}{(2x)^{k}},
\]
where $(n,k):=(-1)^{k}(n+1/2)_k(-n+1/2)_k/k!$ and $a_k$ is the shifted factorial: $a_0=1, a_k = a(a+1)\dots (a+k-1)$ for an integer $k>0$.
The $\pm$ sign must be taken $+$ for $-\pi /2 < \arg x < 3 \pi /2$ and $-$ for $- 3 \pi /2 < arg x < \pi/2$.}.

In this way we get the large $\nu$ expansion for a single Wilson loop 
\begin{equation}
\begin{split}
W_{0}(x=1)_{\nu \to \infty} = 
e^{2 \sqrt{\nu }} \left(
\frac{1}{2 \sqrt{\pi } \nu ^{3/4}}-
\frac{3}{32 \sqrt{\pi } \nu^{5/4}}
-\frac{15}{1024 \sqrt{\pi } \nu ^{7/4} }+\dots
\right) \\
+ e^{-2 \sqrt{\nu }} \left(-\frac{i}{2 \sqrt{\pi } \nu ^{3/4}}-\frac{3 i}{32 \sqrt{\pi } \nu ^{5/4}}+\frac{15 i}{1024 \sqrt{\pi }
   \nu ^{7/4}} + \dots \right)\,.
\end{split}
\end{equation}
It is remarkable, as first pointed out in \cite{Drukker:2006ga}, that the two saddle points corresponding to $e^{\pm 2\sqrt{\nu}}=e^{\pm \sqrt{\tilde{\lambda}}}$ have a bulk interpretation as stable and unstable supersymmetric worldsheets with equal and opposite regularized areas, as discussed in Section \ref{unstable}.

Now we consider the two-point connected correlator. It is not trivial to make inverse Laplace transform of
the function $G_0(z_1,z_2)$ in (\ref{eq:2-point}) because it does not factorize. However, one can make a trick
differentiating $G_0(z_1,z_2)$ over the parameter $\nu$ and then integrating back \cite{Akemann:2001st}. Indeed,
\begin{equation}
  \label{eq:derivative}
  \partial_{\nu} G_0(z_1,z_2) = \frac{4 \nu +z_1 z_2}{\left(z_1^2-4 \nu \right){}^{3/2} \left(z_2^2-4 \nu \right){}^{3/2}}
\end{equation}
we see that $\p_{\nu} G_0(z_1,z_2)$ as a sum of two factorisable terms. Making inverse Laplace transforms of the factors in~(\ref{eq:derivative})
\begin{equation}
  \label{eq:laplace-trans-parts}
\begin{aligned}
   \frac{1}{ (z^2 - 4\nu)^{3/2}} \mapsto \frac { x I_{1} (2 \sqrt{\nu} x)}{2 \sqrt{\nu}} \\
  \frac{z}{ (z^2 - 4\nu)^{3/2}} \mapsto x I_{0} (2 \sqrt{\nu} x)\,,
\end{aligned}
\end{equation}
we get
\begin{equation}
  \label{eq:for-partial}
 \partial_{\nu} G_0 (z_1, z_2) \mapsto \partial_{\nu} W_0(z_1,z_2) = x_1 x_2 (I_1(2 \sqrt{\nu} x_1) I_2(2\sqrt{\nu}x_2) + I_0(2 \sqrt{\nu} x_1) I_0 (2\sqrt{\nu} x_2))\,.  
\end{equation}
After integration over $\nu$ with boundary condition $W_0(z_1,z_2)|_{\nu = 0} = 0$ we get
\begin{equation}
  \label{eq:W-result}
  W_{0}(x_1,x_2) = \frac{\sqrt{\nu }  x_1 x_2  \left( I_0\left(2 \sqrt{\nu } x_1\right)
   I_1\left(2 \sqrt{\nu } x_2\right) + I_0\left(2 \sqrt{\nu } x_2\right) I_1\left(2 \sqrt{\nu } x_1\right)
\right)}{x_1+x_2}
\end{equation}
To evaluate the correlator of coincident Wilson loops with opposite orientation we need to take the limit $x_1 = -x_2 = 1$ of the above
expression\footnote{The choice of same orientations corresponds to $x_1=x_2=1$. Explicit expressions for this case can be found e.g. in \cite{Akemann:2001st}.}. We get
\begin{equation}
  \label{eq:after-limit}
  W_0(x_1,x_2)|_{x_1=-x_2=x} = -x^2 \nu  \left(I_0\left(2 x \sqrt{\nu }\right){}^2+I_2\left(2 x \sqrt{\nu }\right) I_0\left(2 x \sqrt{\nu }\right)-2 I_1\left(2
   x \sqrt{\nu }\right){}^2\right)\,. 
\end{equation}

The large $\nu$ asymptotic expansion gives
\begin{equation}
  \label{eq:expansion-large-nu}
\begin{split}
  W_0(x_1,x_2)|_{x_1 = -x_2 = 1, \nu \to \infty} = 
e^{4 \sqrt{\nu }} \left(-\frac{1}{16 \pi  \sqrt{\nu }}-\frac{3}{128 \pi  \nu }-\frac{33}{2048 \pi  \nu ^{3/2}}+ \dots \right  ) 
\\ 
+e^{-4
   \sqrt{\nu }} \left(\frac{1}{16 \pi  \sqrt{\nu }}-\frac{3}{128 \pi  \nu }+\frac{33}{2048 \pi  \nu ^{3/2}}+\dots \right )\\ 
+\left(-\frac{2 i
   \sqrt{\nu }}{\pi }-\frac{i}{16 \pi  \sqrt{\nu }}-\frac{3 i}{1024 \pi  \nu ^{3/2}} + \dots \right)\,.
\end{split}
\end{equation}

Notice that there are three contribution to the series corresponding to the exponents of $4 \sqrt{\nu}, -4 \sqrt{\nu}$ and $0$. Indeed, as explained in Section \ref{coincident-lat} in the case in which the loop is a latitude on $S^2$, these contributions can be observed in the dual type IIB string theory in the $AdS_5 \times S^5$ background. They correspond to string solutions with regularized areas equal to $S_{reg}=\pm 2\sqrt{\tilde{\lambda}}$ and $S_{reg}=0$, see eq. (\ref{saddle-connected})-(\ref{saddles-coinc}).

We also quote the small $\nu$ expansion of the connected correlator for reference
\begin{equation}
  \label{eq:small-nu}
  W_0(x_1,x_2)|_{x_1 = -x_2 = 1,\nu \to 0 } = 
-\nu - \frac 1 2 \nu^2 - \frac 1 6 \nu^3 - \frac 5 {144} \nu^4 - \frac {7} {1440} \nu^5 + \dots
\end{equation}

\subsection{The connected correlator of two distinct Wilson loops}
\label{2MM-section}


Arguments based on the localization to the two-dimensional
theory \cite{Drukker:2007qr,Drukker:2007yx} and
\cite{Pestun:2007rz,Pestun:2008thesis,Pestun:2009nn} lead us to the following 
$K$-matrix model to compute correlators of $K$ non-self-intersecting
Wilson loops which have topology of $K$ latitudes on $S^2$
\begin{equation}
  \label{eq:K-matrix-model}
\begin{split}
  \la W_{R_1}({\cal C}_1) \dots W_{R_K}({\cal C}_K) \ra
=\frac 1 Z \int_{I \g} [d \phi_1] \dots [d \phi_K] \\
e^{ -\frac {A}{2g_{YM}^2} \tr ( \frac 1 {A_1} \phi_1^2  
+ \frac {1} {A_{12}} (\phi_2 - \phi_1)^2 + \dots 
+ \frac {1} {A_{k,k+1}}(\phi_{k+1}-\phi_{k})^2 + \dots
+ \frac {1} {A_{K}}(\phi_{K})^2 ) } \times \\ \times 
\tr_{R_1} e^{\phi_1} \dots \tr_{R_k} e^{\phi_k} 
\dots \tr_{R_K} e^{\phi_K},
\end{split}
\end{equation}
where $g_{YM}^2$ is the coupling constant of the 4d $\CalN=4$ SYM, 
$A_1$ is the area enclosed by the first loop,  $A_{k,k+1}$ are the areas between
the $k$-th and the $k+1$-th Wilson, and $A_{K}$ is the area enclosed by
the last loop. There are $K+1$-terms in this matrix model corresponding
to the $K+1$ effective propagators on the cylinders between Wilson loops. 
The first and the last propagator is contracted with the Dirac delta like
wave-function corresponding to the the vacuum states 
on the North and the South poles of the sphere.\footnote{
A ``quick and dirty'' way to get the matrix model (\ref{eq:K-matrix-model}) is to recall
that \cite{Blau:1991mp,Witten:1992xu,Blau:1993hj} 2d YM on Riemann
surface $\Sigma$ localizes on classical configurations: $d_{A} *F = 0$. Also, 
in two dimensions, Yang-Mills theory  essentially reduces to the abelian theory~\cite{Blau:1991mp,Blau:1993hj},
so in our case we can consider $*F$ to be a piecewise constant function on $\Sigma$ 
with jumps supported on Wilson loops. The contribution 
to the 2d YM action coming from $\Sigma_{k,k+1}$ (the region between $k$-th and $(k+1)$-th Wilson loop) is 
\begin{equation}
  \label{eq:2d-YM-contribution}
  S_{k,k+1} = -\frac {1} {g_{2d}^2} \int_{\Sigma_{k,k+1}} d^2 x \sqrt{g} \tr (*F)^2  = - \frac {1}{g_{2d}^2}A_{k,k+1} \tr X_{k,k+1}^2\,,
\end{equation}
where $X_{k,k+1}$ denotes $*F$ on $\Sigma_{k,k+1}$. Using Stokes theorem on $\Sigma_{k,k+1}$ 
\begin{equation}
  \label{eq:Stokes-2d}
 A_{k,k+1} X_{k,k+1} =  \int_{\Sigma_{k,k+1}} d^2 x \sqrt{g} *F  = \int_{\Sigma_{k,k+1}} F = \int_{C_{k+1}} A - \int_{C_{k}} A = \phi_{k+1} - \phi_{k}\,,
\end{equation}
we get (\ref{eq:K-matrix-model}) and  the relation $g_{2d}^2 = - \frac {2 g_{YM}^2}{A}$. In this derivation we assumed that after the abelianization the gauge bundle is trivial, and this precisely means dropping out the unstable instantons \cite{Blau:1991mp}\cite{Blau:1993hj}.}

To be more concrete, recall that the two-dimensional Yang-Mills on a cylinder
with the action 
\begin{equation}
  \label{eq:2d-ym}
 S_{2d} = - \frac {1} {2 g_{2d}^2} \int d^2 x \sqrt{g} \tr F_{\alpha \beta} F^{\alpha \beta}  
\end{equation}
 in Hamiltonian formalism is equivalent to the quantum mechanics on the group manifold $G$ 
with the Hamiltonian\footnote{In Euclidean space
our convention is $\frac{\p \Psi}{\p t} = -H \Psi$ where $\Psi$ is the wave-function.}
\begin{equation}
  \label{eq:hamitonian-2d-YM}
  -H = \frac{g_{2d}^2} {4} \Delta = \frac {g_{2d}^2}{4} C_{2}\,, 
\end{equation}
where $\Delta$ is the Laplacian on $G$, the $C_2$ is the second Casimir operator\footnote{In our conventions the
basis $T_a$ on $\g$ is normalized as
 $\tr_F T_{a} T_{b} = -\frac 1 2 \delta_{ab}$ and 
we choose the metric on $\g$ in this basis to be $\frac{1}{2} \delta_{ab}$.
The Laplacian for this metric (Casimir operator) is then $\Delta =C_2 = \sum e_a e^a = 2 \sum_{a} e_a e_a$, 
where $e_a$ are basis elements in $\g$ corresponding to $T_a$. The constant $C_2(R)$ 
is defined as $T_a(R) T_a(R) = -C_2(R) {\mathbf 1}$. Hence $\Delta \Psi_{R} = - 2 C_2(R)\Psi_{R}$ if $\Psi_{R}$ is a matrix element of $G$ in representation~$R$.}, 
and the time is measured in the units of area thanks to the usual area-preserving diffeomorphism invariance of the 2d Yang-Mills,
see e.g. \cite{Migdal:1975zf, Kazakov:1980zi, Kazakov:1981sb, Rusakov:1990rs, Blau:1993hj,Blau:1991mp,
Gross:1989fc,Fine:1990zz,Witten:1991we,Gorsky:1993pe,Gross:1994ub}.

The configurational space here is the holonomy of the gauge connection around a spacial circle, i.e. it coincides with the gauge group $G$. 
The wave-function  $\psi$ is a function on $G$, i.e. we write $\psi(U)$ for $U \in G$.
The heat kernel $K(U,1;t)$ for the Laplacian (\ref{eq:hamitonian-2d-YM})
 defines the wave-function of the theory at time $t$ starting from the Dirac-delta function supported at 
the group unity at time $t=0$ \cite{Migdal:1975zf,Witten:1991we}
 \begin{equation}
   \label{eq:heat-kernel}
   K(U,1;t) = \sum_{R} d_{R} \exp (-\frac {g_{2d}^2 t }{2} C_{2} (R)) \chi_{R} (U)\,,
 \end{equation}
where $\chi_{R}$ is the character of $U \in G$ in representation $R$ and the sum is taken 
over all irreducible representations of $G$.\footnote{We remind that the characters for a compact simple group $G$ are orthonormal
\[ \frac 1 {\vol{G}} \int_{G} [dU] \chi_{R} (U) \bar \chi_{R'} (U) = \delta_{R R'} \]
and that the delta-function at the group unity is expressed as
\[ \delta(U-1) = \sum_{R} d_{R} \chi_{R}(U)\]
where the sum is taken over all irreducible representations of $G$.
} 
More generally the propagator between two adjoint invariant
states represented by $U_1$ and $U_2$ is 
\begin{equation}
  \label{eq:heat-kernel-2prop}
  K(U_1,U_2;t) = \sum_{R} \chi_{R}(U_1) \bar \chi_{R}(U_2) \exp ( - \frac {g_{2d}^2 t }{2} C_{2} (R)).
\end{equation}

The expectation value of $K$ Wilson loops on a cylinder located 
at times $A_1, A_1+A_{12}, \dots$ and with trivial initial and the final state 
is then 
\begin{equation}
  \label{eq:exact-2d}
\begin{split}
  \langle W_{R_1} \dots W_{R_K} \rangle =  \\
\int [dU_1] \dots [dU_K] \, K(1,U_1;A_1) \, K(U_1,U_2;A_{12}) \dots K(U_K,1;A_{K}) \times \\ 
\tr_{R_1} U_1 \tr_{R_2} U_2 \dots
\tr_{R_K} U_K\,.
\end{split}
\end{equation}
The integral in (\ref{eq:exact-2d}) is taken over $K$ copies of the group manifold $G$.

However, we can relate this integral to the integral over the Lie algebra $\g$ of $G$ up to 
the exponentially small instanton corrections \cite{Bassetto:2001mf}. 
We consider $G=\U(N)$ gauge group for brevity, but the same arguments are applicable to any simple compact Lie group.
Let $\theta_{i}$ be coordinates on the Cartan subalgebra $\h$ of $\g$, and let $\diag(e^{i\theta_1}, 
\dots, e^{i \theta_N})$ be the image of $(\theta_1,\dots, \theta_N) \in \h$ 
under the exponential map $\g \to G$. Clearly the integral in (\ref{eq:exact-2d}) is invariant
under adjoint transformations of $U_k$ for $k=1\dots K$, and hence we can reduce it to the integral over diagonal 
unitary matrices. Let $\theta_i$ for $i = 1 \dots N$ be coordinates 
on the maximal torus $T \in G$ with $T$ being the image of $\h$ under the exponential map,
and let $t_{ij}$ for $i \neq j $ be local coordinates on $G$ in a neighborhood of $T$. 
Let $T_{ij}, i \neq j$ be generators of $\g / \h$, so that $t_{ij} T_{ij} \in \g / \h$ parametrize the normal directions to $T$. One can take
$\Lambda = \diag (e^{i\theta_1}, \dots, e^{i\theta_N})$, then $W=e^{t_{ij} T_{ij}}$ and
\begin{equation}
  \label{eq:diag}
  U = W \Lambda W^{-1}\,.
\end{equation}
In these coordinates the metric on $G$ in the neighboorhood of $T$ has the familiar form 
\begin{equation}
  \label{eq:metric}
  ds^2 = \sum_{i=1}^{N} d \theta_i^2 + \sum_{1 \leq i \neq j \leq N} (2 \sin ( \frac{ \theta_j - \theta_i} 2))^2 dt_{ij}^2\,. 
\end{equation}
Using this metric and the standard formula for the Laplacian
\begin{equation}
  \label{eq:Laplicain}
  \Delta_G = \frac {1} {\sqrt{g}} \p_{\mu} (\sqrt {g} g^{\mu \nu} \p_{\nu})\,,
\end{equation}
we can find that on the adjoint invariant functions on $G$ represented as functions on $T$ and denoted as $K(\theta_1,\dots, \theta_N)$ 
the Laplacian acts as
\begin{equation}
  \Delta_G K = \sum_{i=1}^{N} \frac {1} {J^2} \frac {\p} {\p \theta_{i}} \lb J^2 \frac {\p} {\p \theta_i} K \rb\,, 
\end{equation}
where 
\begin{equation}
  \label{eq:J-def}
  J(\theta) = \prod_{i < j} (2 \sin ( \frac{ \theta_j - \theta_i} 2))\,.
\end{equation}

Now notice that 
\begin{equation}
  \label{eq:fun}
  \sum_{i} \frac {1} {J^2} \frac {\p} {\p \theta_{i}} \lb J^2 \frac {\p} {\p \theta_i} f \rb  =
  \frac 1 J \frac {\p^2} {\p \theta^2} \lb J f \rb -   \lb \frac 1 J \frac {\p^2} {\p \theta^2} J \rb f \,.
\end{equation}

It is a very nice algebraic identity which can be proved using Weyl formula that the last term is
a constant
\begin{equation}
  \label{eq:idenitity}
  \frac 1 J \frac {\p^2} {\p \theta^2} J(\theta) = - \frac {1} {12} N (N^2-1),
\end{equation}
hence we get 
\begin{equation}
  \label{eq:laplacian-J}
  \Delta_G K = \frac 1 J \frac {\p^2} {\p \theta^2} \lb J K \rb +  \frac {1} {12} N (N^2-1) K\,.
\end{equation}
The addition of a constant to the Hamiltonian 
only shifts the zero energy level and does not change the correlation functions, 
so for the purpose of computing correlation functions of Wilson loop operators we can drop it.
Then we end up with a simple formula 
\begin{equation}
  \label{eq:laplacian-J1}
  \tilde \Delta_G K = \frac 1 J \frac {\p^2} {\p \theta^2} \lb J K \rb\,.
\end{equation}

Repeating the same arguments for the Lie algebra $\g$ instead of $G$ we can get Laplacian on 
the adjoint invariant functions on $\g$ 
\begin{equation}
  \label{eq:laplacian-on-g}
  \Delta_\g K= \frac 1 J_0 \frac {\p^2} {\p \theta^2} \lb J_0 K \rb \,,
\end{equation}
where
\begin{equation}
  J_0(\theta) = \prod_{i < j} (\theta_j - \theta_i)\,. 
\end{equation}

Looking at  the Laplacian (\ref{eq:laplacian-J}) on $G$ and the Laplacian (\ref{eq:laplacian-on-g}) we
notice that heat kernel equation for the functions $\tilde K_{G} = J K_{G}$ and $\tilde K_{\g} = J_0 K_{\g}$  is the
same usual heat kernel equation in the flat space $R^{N}$
\begin{equation}
  \label{eq:canonical}
  \frac {\p} {\p t} \tilde K  = \frac {g^2_{2d}}{4} \sum_{i=1}^{N} \frac { \p^2} {\p \theta^2_i} \tilde K\,.
\end{equation}

However, strictly speaking, the solution $\tilde K_{G}$ on $G$ and $\tilde K_{\g}$ on $\g$ is not
the same. The difference is the global structure of the space. The exponential map $\g \to G$ 
is a local isomorphism between $\g$ and $G$ but, of course, not a global
one. In particular, the points on the integer lattice $(2 \pi n_1, \dots
, 2\pi n_k)$ in $\h$ are mapped to the unity of the $G$. Hence, while $\tilde K_{\g}$ satisfies the heat kernel
equation (\ref{eq:canonical}) with a single delta-function source inserted at the point $\theta_i = 0$,
the $\tilde K_{G}$ satisfies the same heat kernel equation (\ref{eq:canonical}) but 
with infinitely many images of the delta-function source inserted at the points on the 
lattice $(2 \pi n_1, \dots , 2\pi n_k), n_1,\dots, n_k \in \BZ$. The
contribution of the images for $\vec{n} \neq 0$ is exponentially
suppressed for $g_{2d} \to 0$ and corresponds to the instanton
contribution to the 2d YM partition function.

Rewriting the unitary matrix model (\ref{eq:exact-2d}) as integral over eigenvalues $\vec{\theta}_k$ 
for $k = 1 \dots K$ we get (here $U_1=U_1(\vec\theta_1) \dots U_K=U_K(\vec \theta_K)$)
\begin{equation} 
\begin{split}
  \langle W_{R_1} \dots W_{R_K} \rangle =  \\
\int [d^N \vec \theta] \dots [d^N \vec \theta] \, K_G(1,\vec \theta_1 ;A_1) 
J(\vec \theta_1) J( \vec \theta_1) \, K_G(\vec \theta_1, \vec \theta_2;A_{12}) J(\vec \theta_2)^2\dots 
K_G(\vec \theta_K,1;A_{K}) \times \\ 
\chi_R(\vec \theta_1)  \dots \chi_R (\vec \theta_K)\,.
\end{split}
\end{equation}

We argued that $ J(\vec \theta_1) K_{G} (\vec \theta_1, \vec \theta_2, t) J(\vec \theta_2)$
is equal to $ J_0 ( \vec \theta_1) K_{\g} (\vec \theta_2, \vec \theta_2, t) J_0(\vec \theta_2)$
up to the instanton corrections. In other words, the heat kernel $K_{G}$ on invariant
functions on $G$ multiplied the measure factors $J(\vec \theta)$ which appear in the reduction 
of the unitary matrix integrals to the eigenvalues, is equal, up to instanton corrections, to 
the heat kernel $K_{g}$ on invariant functions on $\g$ similarly multiplied by the factors $J_{0}$
which appear in the reduction of the Hermitian matrix integral to the eigenvalues. Hence the ``zero-instanton" contribution to the correlator is 
\begin{equation} 
\label{eq:2d-ym-matrix-model}
\begin{split}
  \langle W_{R_1} \dots W_{R_K} \rangle_{0-inst} =  \\
\int [d^N \vec \theta] \dots [d^N \vec \theta] \, K_\g(1,\vec \theta_1 ;A_1) 
J_0(\vec \theta_1) J_0( \vec \theta_1) \, K_\g(\vec \theta_1, \vec \theta_2;A_{12}) J_0(\vec \theta_2)^2\dots 
K_\g(\vec \theta_K,1;A_{K}) \times \\ 
\chi_R(\vec \theta_1)  \dots \chi_R (\vec \theta_K)\,. 
\end{split}
\end{equation}

Explicitly, for generic $\vec \theta_1$ and $\vec \theta_2$ we have
\begin{equation}
  \label{eq:K-propagator}
  J_0(\vec \theta_1) K_{\g}(\vec \theta_1, \vec \theta_2, t) J_0(\vec
  \theta_2) = \const \frac {1} {t^{N/2}} \sum_{\sigma \in S_N}
  (-1)^{\sigma} \exp \lb - \frac {1} { g_{2d}^2 t } { (\vec \theta_1 -
    \sigma( \vec \theta_2))^2  } \rb\,,
\end{equation}
where $S_N$ is the permutation group of $N$ elements. For a general Lie
algebra the permutation group $S_N$ should be replaced by the Weyl group. It is
clear that the right hand side of (\ref{eq:K-propagator}) satisfies the
usual heat equation on the flat space $\BR^{N}$ (\ref{eq:canonical}).
Also, when we reduce the integral over the Lie algebra to its Cartan subalgebra there is 
still a discrete residue symmetry group left -- the Weyl group. 
The summation over Weyl group in (\ref{eq:K-propagator}) projects to the Weyl invariant state. 
The heat kernel  $K_{\g}(\vec \theta_1, \vec \theta_2, t)$ is a positive function and is Weyl 
invariant (completely symmetric in $\theta_{1i}$ or $\theta_{2i}$), while  
 $J_0(\vec \theta_1) K_{\g}(\vec \theta_1, \vec \theta_2, t) J_0(\vec  \theta_2)$
is Weyl skew-invariant. 

If the source is the origin of the Lie algebra, i.e. at $\theta_i = 0$, then 
the solution (\ref{eq:K-propagator}) is degenerate and is not applicable. 
The correct solution is 
\begin{equation}
  \label{eq:K-propagator1}
   K_{\g}(1, \vec \theta_2, t) J_0(\vec
  \theta_2) = \const \frac {1} {t^{N^2/2}} 
 J_0(\vec \theta_2) \exp \lb - \frac {1} {  g_{2d}^2 t } { (\vec \theta_2)^2 } \rb\,.
\end{equation}
For more details see e.g. \cite{Bassetto:2001mf} and references therein.

Replacing the integral (\ref{eq:2d-ym-matrix-model}) over the eigenvalues of the Hermitian matrices by the integral over all 
Hermitian matrices we recover the Hermitian $K$-matrix model chain (\ref{eq:K-matrix-model}) 
up to a factor of $i=\sqrt{-1}$ in the definition of observables. In other words, 
the Wilson loops in the 2d YM with compact gauge group $G$ are unitary, e.g. in the fundamental 
representation $\chi_{f} (\vec{\theta}) =\sum_{j=1}^{N} \exp( 2 \pi i \theta_j)$ with $\theta_j$ being real. 
The observables in the Hermitian matrix model which describe the supersymmetric Wilson loops of the $d=4$ $\CalN=4$ Yang-Mills
are not unitary, but Hermitian, i.e. the fundamental Wilson loops are represented in the Hermitian matrix 
integral as $\chi_{f} (\vec{\theta}) =\sum_{j=1}^{N} \exp( 2 \pi  \theta_j)$ with the same convention that $\theta_j$ are real.

One can equivalently formally represent $d=4$ $\CalN=4$ Yang-Mills SUSY Wilson loops by 
the usual unitary Wilson loop in the bosonic $2d$ Yang-Mills in the zero instanton sector 
and with the inverted sign of the 2d YM coupling constant 
\begin{equation}
  \label{eq:relation}
  g_{2d}^2 = - \frac { 2 g_{4d}^2} {A}
\end{equation}
($A= 4 \pi r^2$ is the total area of the $S^2$).

In the case of two Wilson loops we can write the two-matrix model action
in the form
\begin{equation}
  \label{eq:two-matrix-model}
Z_{2MM} = \int [d\phi_1] [d\phi_2] e^{ -N \tr (V_1 (\phi_1) + V_2 (\phi_2) - c \phi_1 \phi_2)}\,,
\end{equation}
with two quadratic potentials $V_1(\phi_1) = \frac 1 2 a_1 \phi_1^2, 
V_2(\phi_2))=\frac 1 2 a_2 \phi_2^2$ and
the interaction coupling $c$. In terms of the original areas
and the 't Hooft coupling constant $\lambda = g_{YM}^2 N$ we have
\begin{equation}
  \label{eq:a1-a2-c-expression}
\begin{aligned}
a_1&=\frac {A} {\lambda} \lb \frac {1} {A_1} + \frac {1} {A_{12}} \rb \\
a_2&=\frac {A} {\lambda} \lb \frac {1} {A_2} + \frac {1} {A_{12}} \rb \\
c  &=\frac {A} {\lambda} \frac {1} {A_{12}}.
\end{aligned}
\end{equation}

We now solve in the planar limit the correlator (\ref{eq:K-matrix-model}) in the case of the two-matrix model and for choice of fundamental representation. Following \cite{eynard-1997}, define two resolvent functions
\begin{align}
   \label{eq:resolvent-function1}
   z_1 &= \alpha x + \alpha_1 \frac 1 x \\
   \label{eq:resolvent-function2}
   z_2 &= \alpha \frac 1 x + \alpha_2 x \,,
\end{align}
where $\alpha,\alpha_1,\alpha_2$ are coefficients determined from the
algebraic equations
\begin{equation}
  \label{eq:algebraic-equations-to-find-alpha}
\begin{aligned}
  \frac 1 c V_1'(z_1) - z_2 = \frac 1 c  \frac 1 \alpha \frac 1 x \\
  \frac 1 c V_2'(z_2) - z_1 = \frac 1 c  \frac 1 \alpha x\,.
\end{aligned}
\end{equation}

Solving these equations on $\alpha,\alpha_1,\alpha_2$ we find
\begin{equation}
  \label{eq:alpha-solution}
  \begin{aligned}
    \alpha^2 = \frac {c} {a_1 a_2 - c^2} \\
    \alpha_1 = \frac {a_2} {c} \alpha \\
    \alpha_2 = \frac {a_1} {c} \alpha\,.
  \end{aligned}
\end{equation}

Now, inverting the equation (\ref{eq:resolvent-function1})
we get the resolvent function $x_1(z_1)$ and inverting
the equation (\ref{eq:resolvent-function2}) we get
the resolvent function $x_2(z_2)$
\begin{equation}
  \label{eq:x1-x2-solutions}
\begin{aligned}
 &x_1=\frac {1} {2 \alpha} \lb z_1  \pm \sqrt{z_1^2 - 4 \alpha \alpha_1} \rb\\
 &x_2=\frac {1} {2 \alpha_2}\lb z_2 \pm \sqrt{z_2^2 -4 \alpha \alpha_2}\rb\,. 
\end{aligned}
\end{equation}

In \cite{eynard-1997} and \cite{Daul:1993bg} it was shown that the planar limit
of the two-point resolvent function
for the matrices $\phi_1,\phi_2$
\begin{equation}
  \label{eq:resolvent-function-def}
  \omega(z_1,z_2) = \la \tr \frac {1} {z_1 - \phi_1} 
  \tr \frac {1} {z_2 - \phi_2} \ra_{\rm conn}
\end{equation}
is given by  
\begin{equation}
  \label{eq:resolvent-Eynard-result}
  \omega(z_1,z_2) = -\frac {\p} {\p z_1} \frac {\p} {\p z_2} 
\log \left (1 - \frac {x_1(z_1)} {x_2(z_2)} \right)\,.
\end{equation}
To cast this formula in more symmetrical form with respect 
to the two loops we will denote 
\begin{equation}
  \label{eq:y-notations}
  y_1 (z_1) \equiv x_1 (z_1) \quad y_2(z_2) \equiv \frac 1 {x_2(z_2)}
\end{equation}
so 
\begin{equation}
  \label{eq:resolvent-Eynard-result-y-not}
  \omega(z_1,z_2) = -\p_{z_1} \p_{z_2} \log \left (1 - y_1(z_1) y_2(z_2) \right)\,,
\end{equation}
where now 
\begin{equation}
  \label{eq:y-explicit}
\begin{aligned}
y_1 (z_1) = \frac 1 {2 \alpha} \lb z_1 + \sqrt{z_1^2 - 4 \alpha \alpha_1} \rb \\
y_2 (z_2) = \frac 1 {2 \alpha} \lb z_2 + \sqrt{z_2^2 - 4 \alpha \alpha_2} \rb \,.
\end{aligned}
\end{equation}

To compute correlators of Wilson loops we need the loop  generating function
\begin{equation}
  \label{eq:w-generating}
  W(t_1,t_2) = \la \tr e^{t_1 \phi_1} \tr e^{t_2 \phi_2} \ra_\conn\,.
\end{equation}

The function $W(t_1,t_2)$ is obtained from $\omega(z_1,z_2)$ by the 
inverse Laplace transform
\begin{equation}
  \label{eq:inverse-Laplace-2mm}
  W(t_1,t_2) = \frac {1} {(2\pi i)^2} \oint dz_1 dz_2 \omega(z_1,z_2) 
e^{z_1 t_1 + z_2 t_2},
\end{equation}
where the  integration contours are sufficiently large to encircle the
cuts in the $z_1$ and the $z_2$ plane.

The choice of signs in (\ref{eq:y-explicit}) and the integration contours is fixed by comparing the resolvent 
with the leading order Feynman diagrams computed directly from the matrix model (\ref{eq:two-matrix-model}).

\subsubsection{Large $\lambda$ asymptotic} 

Now we will compute (\ref{eq:inverse-Laplace-2mm}) by the stationary phase
method in the limit $\lambda \to \infty$. 
From (\ref{eq:resolvent-Eynard-result}) we have 
\begin{equation}
  \label{eq:omega-in-y}
  \omega(z_1,z_2) = \frac { \p_{z_1} y_1 (z_1) \p_{z_2} y_2 (z_2) }
{ (1 - y_1 (z_1) y_2 (z_2) )^2 }\,. 
\end{equation}

It is convenient first to change integration variables in
the integral (\ref{eq:inverse-Laplace-2mm}) from $z_i$ to $y_i$ 
\begin{equation}
  \label{eq:y-z-relation}
  \begin{aligned}
  z_i = \alpha y_i + \frac { \alpha_i} { y_i } \quad i = 1,2\,.
  \end{aligned}
\end{equation}
So we get 
\begin{equation}
  \label{eq:W(t1,t2)(y)}
  W(t_1,t_2) = \frac {1} { (2 \pi i)^2 } \oint \frac {dy_1 dy_2 } {
    (1-y_1 y_2)^2 } e^{ t_1 (a y_1 + \frac {\alpha_1} {y_1})} 
e^{ t_2(a y_2 + \frac {\alpha_2} {y_2})}\,.
\end{equation}
We can integrate over the circular contours of radius $r_i > 1$
\begin{equation}
  \label{eq:contours}
  y_i = r_i e^{i \phi_i}, \quad r_i = \sqrt { \frac {\alpha_i}{\alpha}
  },
\quad \phi_i \in [0, 2\pi)
\end{equation}
so 
\begin{equation}
  \label{eq:circular-contours}
  W(t_1,t_2) = \frac {1} {(2\pi)^2} \int d \phi_1\, d \phi_2 
\frac { r_1 r_2 e^{i \phi_1 + i \phi_2} } { (1 - r_1 r_2 e^{(i\phi_1 + i
    \phi_2)} )^2 } e^{2 t_1 \sqrt{\alpha \alpha_1} \cos \phi_1} 
e^{2 t_2 \sqrt{ \alpha \alpha_2} \cos \phi_2}\,.
\end{equation}

Notice that for not coincident loops $ r_1 r_2 > 1$ (see
(\ref{eq:subst-alpha})) so we replace the integrand by the
convergent series in inverse powers of $(r_1 r_2)$ 
\begin{equation}
  \label{eq:series}
  \frac {y_1 y_2} { (1 - y_1 y_2)^2 } = \sum_{n=1}^{\infty} n (y_1 y_2)^{-n}\,.
\end{equation}
Each term in the series factorizes into a simple product of Bessel
functions. Using (\ref{eq:integral-repre}) we get 
\begin{equation}
  \label{eq:W-good-series}
  W(t_1,t_2) = \sum_{n=1}^{\infty} n (r_1 r_2)^{-n} 
I_n(2 t_1 \sqrt{ \alpha \alpha_1}) I_n(2 t_2 \sqrt{\alpha \alpha_2})\,. 
\end{equation}

Let us denote 
\begin{equation}
  \label{eq:notations}
  \rho = r_1 r_2 = \sqrt{\frac {\alpha_1 \alpha_2} { \alpha^2}}  \quad \tilde t_1 = 2 t_1 \sqrt{\alpha \alpha_1} 
  \quad \tilde t_2 = - 2 t_2 \sqrt{ \alpha \alpha_2}\,. 
\end{equation}
We introduced the minus sign in the definition of $\tilde t_2$ for convenience in order to specialize to the case of oppositely oriented Wilson loops for which $t_1 > 0, t_2 < 0$, so both $\tilde t_1, \tilde t_2$ are positive. 
Then 
\begin{equation}
  \label{eq:W-good-series-t-2-rewrite}
\begin{split}
  W(\tilde t_1, \tilde t_2) = \sum_{n=1}^{\infty} n \rho^{-n} (-1)^n I_n(\tilde t_1) I_n(\tilde t_2) = \\
= \sum_{n=1}^{\infty} n \rho^{-n} (-1)^{n} \frac {1} {\sqrt{2 \pi \tilde t_1}} \frac {1} {\sqrt{ 2 \pi \tilde t_2}}
\left( e^{\tilde t_1} + i (-1)^n e^{-\tilde t_1} + \dots \right ) 
\left( e^{\tilde t_2} + i (-1)^n e^{-\tilde t_2} + \dots \right ) = \\
= \frac {1} {\sqrt{2 \pi \tilde t_1}} \frac {1} {\sqrt{ 2 \pi \tilde t_2}} \left ( \frac { - \rho} { (1+ \rho)^2} \left ( e^{\tilde t_1 + \tilde t_2} - e^{-\tilde t_2  - \tilde t_1} \right)
+ i \frac { \rho} { (1 - \rho)^2} \left ( e^{ -\tilde t_1 + \tilde t_2} + e^{\tilde t_1 - \tilde t_2} \right) + \dots \right )\,.
\end{split} 
\end{equation}


The result in terms of the areas $A_1,A_2,A_{12}$ and 't Hooft coupling
constant $\lambda$ is easily obtained using (\ref{eq:notations}) at $t_1 = -t_2 = 1$
and (\ref{eq:a1-a2-c-expression}), (\ref{eq:alpha-solution}), which give
\begin{equation}
  \label{eq:subst-alpha}
\begin{aligned}
&\alpha = \sqrt{ \frac { \lambda A_1 A_2 } {A^2}} \\
&\tilde t_1 =  2  \sqrt{\alpha \alpha_1} =  2 \sqrt{ \frac { \lambda} {A} \frac{
      (A-A_1)A_1}{A}}\,,\qquad 
\tilde t_2  = 2   \sqrt{\alpha \alpha_2} = 2 \sqrt{ \frac { \lambda} {A} \frac{ (A-A_2)
    A_2} {A}} \\
&\rho = \sqrt{\frac {\alpha_1 \alpha_2}{\alpha^2}} = 
\sqrt{\lb 1 + \frac {A_{12}}{A_{1}} \rb
\lb 1 + \frac {A_{12}}{A_2} \rb}\,.
\end{aligned}
\end{equation}
Therefore, the prediction of the two-matrix model for the strong coupling behavior of the correlator is, at leading order,
\begin{equation}
W(1,-1) \stackrel{\lambda \rightarrow \infty}{\simeq} \exp\left[ \sqrt{\frac{\lambda\,4 A_1(A-A_1)}{A^2}}+\sqrt{\frac{\lambda\,4 A_2(A-A_2)}{A^2}}\right]\,,
\end{equation} 
which matches with the bulk interpretation given in Section \ref{correl-non-coinc} of a singular cylinder made up of two disconnected disks joined by a supergraviton propagator. Moreover, the presence of four saddle points in (\ref{eq:W-good-series-t-2-rewrite}) is also mirrored in the bulk by the four possible pairings of stable/unstable disk solutions. It would be very interesting to test the structure of the prefactors to the exponential saddle points in (\ref{eq:W-good-series-t-2-rewrite}) by computing the supergraviton exchange on the string theory side.



\subsection*{Acknowledgments}
We thank Nadav Drukker, Anatoly Dymarsky, Takuya Okuda and Diego Trancanelli for interesting discussions and useful comments.
The work of S.G. is supported in part by the Center for the
Fundamental Laws of Nature at Harvard University and by NSF grants PHY-024482
and DMS-0244464. The work of V.P. is supported by a Junior Fellowship from the
Harvard Society of Fellows, and grants NSh-3035.2008.2 and RFBR 07-02-00645. The work of R.R. is supported by an EPSRC Fellowship.

\bibliography{bsample}
\end{document}